\documentclass[twocolumn,traditabstract,longauth]{aa}
\usepackage{graphicx}
\usepackage[breaklinks, colorlinks, citecolor=blue]{hyperref}
\usepackage{natbib}
\usepackage{subfigure}
\usepackage{amsmath}
\usepackage{epsfig}
\usepackage{multirow}
\usepackage{color}
\usepackage{txfonts}

\newcommand{\eg}{{e.g.}}
\usepackage{url}
\usepackage{ragged2e}

\def\Msun{\>{\rm M_{\odot}}}
\def\Lsun{\>{\rm L_{\odot}}}

\def\Planck{\textit{Planck}}
\newbox\tablebox    \newdimen\tablewidth
\def\leaderfil{\leaders\hbox to 5pt{\hss.\hss}\hfil}
\def\endPlancktable{\tablewidth=\columnwidth 
    $$\hss\copy\tablebox\hss$$
    \vskip-\lastskip\vskip -2pt}
\def\endPlancktablewide{\tablewidth=\textwidth 
    $$\hss\copy\tablebox\hss$$
    \vskip-\lastskip\vskip -2pt}
\def\tablenote#1 #2\par{\begingroup \parindent=0.8em
    \abovedisplayshortskip=0pt\belowdisplayshortskip=0pt
    \noindent
    $$\hss\vbox{\hsize\tablewidth \hangindent=\parindent \hangafter=1 \noindent
    \hbox to \parindent{$^#1$\hss}\strut#2\strut\par}\hss$$
    \endgroup}
\def\doubleline{\vskip 3pt\hrule \vskip 1.5pt \hrule \vskip 5pt}

\def\WHzsr{\ifmmode $W\,Hz\mo\,sr\mo$\else W\,Hz\mo\,sr\mo\fi}
\def\mHz{\ifmmode $\,mHz$\else \,mHz\fi}
\def\GHz{\ifmmode $\,GHz$\else \,GHz\fi}
\def\mKs{\ifmmode $\,mK\,s$^{1/2}\else \,mK\,s$^{1/2}$\fi}
\def\muKs{\ifmmode \,\mu$K\,s$^{1/2}\else \,$\mu$K\,s$^{1/2}$\fi}
\def\muKRJs{\ifmmode \,\mu$K$_{\rm RJ}$\,s$^{1/2}\else \,$\mu$K$_{\rm RJ}$\,s$^{1/2}$\fi}
\def\muKHz{\ifmmode \,\mu$K\,Hz$^{-1/2}\else \,$\mu$K\,Hz$^{-1/2}$\fi}
\def\MJysr{\ifmmode \,$MJy\,sr\mo$\else \,MJy\,sr\mo\fi}
\def\MJysrmK{\ifmmode \,$MJy\,sr\mo$\,mK$_{\rm CMB}\mo\else \,MJy\,sr\mo\,mK$_{\rm CMB}\mo$\fi}
\def\microns{\ifmmode \,\mu$m$\else \,$\mu$m\fi}
\def\micron{\microns}
\def\muK{\ifmmode \,\mu$K$\else \,$\mu$\hbox{K}\fi}
\def\microK{\ifmmode \,\mu$K$\else \,$\mu$\hbox{K}\fi}
\def\muW{\ifmmode \,\mu$W$\else \,$\mu$\hbox{W}\fi}
\def\kms{\ifmmode $\,km\,s$^{-1}\else \,km\,s$^{-1}$\fi}
\def\kmsMpc{\ifmmode $\,\kms\,Mpc\mo$\else \,\kms\,Mpc\mo\fi}

\providecommand{\sorthelp}[1]{}

\begin{document}
\author{%\small
I.~Flores-Cacho\inst{1,2}
\and
D.~Pierini\inst{1,2,7}
\and
G.~Soucail\inst{1,3}\thanks{\href{mailto:Genevieve.Soucail@irap.omp.eu}{Genevieve.Soucail@irap.omp.eu}}
\and
L.~Montier\inst{1,2}
\and
H.~Dole\inst{4,5}
\and
E.~Pointecouteau\inst{1,2}
\and
R.~Pell{\'o}\inst{1,3}
\and
E.~Le Floc'h\inst{6}
\and
N.~Nesvadba\inst{4}
\and
G.~Lagache\inst{7,4}
\and
D.~Guery\inst{4}
\and 
R.~Ca{\~n}ameras\inst{4}
}
\institute{\small
Universit\'{e} de Toulouse, UPS Toulouse III - OMP, IRAP, Toulouse, France\goodbreak
\and
CNRS, IRAP, Institut de Recherche en Astrophysique et Plan\'etologie, 9 Avenue du Colonel Roche, BP 44346, F-31028 Toulouse cedex 4, France\goodbreak
\and
CNRS, IRAP, 14 Avenue Edouard Belin, F-31400 Toulouse, France\goodbreak
\and
Institut d'Astrophysique Spatiale, CNRS (UMR8617) Universit\'{e} Paris-Sud 11, B\^{a}timent 121, Orsay, France\goodbreak 
\and
Institut Universitaire de France, 103 bd Saint-Michel, F-75005, Paris, France\goodbreak
\and
Laboratoire AIM, IRFU/Service d'Astrophysique - CEA/DSM - CNRS - Universit\'{e} Paris Diderot, B\^{a}t. 709, CEA-Saclay, F-91191 Gif-sur-Yvette Cedex, France\goodbreak
\and
Aix Marseille Universit{\'e}, CNRS, LAM (Laboratoire d'Astrophysique de Marseille) UMR 7326, 38 rue Fr{\'e}d{\'e}ric Joliot-Curie, F-13388 Marseille cedex 13, France\goodbreak
}

\title{Multi-wavelength characterisation of $z\sim2$ clustered, dusty star forming galaxies discovered by \textit{Planck}}

\abstract{
We report the discovery of PHz\,G95.5$-$61.6, a complex structure detected in emission
in the \Planck\ all-sky survey that corresponds to two over-densities of high-redshift (i.e., $z >1$) galaxies.
This is the first source from the \Planck\ catalogue of high-$z$ candidates (proto-clusters and lensed systems)
that has been completely characterised with follow-up observations from the optical
to the sub-millimetre (sub-mm) domain.
{\it Herschel}/SPIRE observations at 250, 350 and 500\microns\ reveal the existence of five sources 
producing a 500\microns\ emission excess that spatially corresponds to the candidate proto-clusters
discovered by \Planck.
Further observations at CFHT in the optical bands ($g$ and $i$), with MegaCam,
and in the near infrared (NIR) ($J$, $H$ and $K_{\rm s}$), with WIRCam,
plus mid infrared (MIR) observations with IRAC/{\it Spitzer}  (at 3.6 and 4.5\,\microns)
confirm that the sub-mm red excess is associated with an over-density of colour-selected galaxies
($i-K_{\rm s}\sim2.3$ and $J-K\sim0.8\,$AB-mag). 
Follow-up spectroscopy of 13 galaxies with VLT/X-Shooter establishes the existence of two high-$z$ structures:
one at $z\simeq1.7$ (three confirmed member galaxies), the other at $z\simeq2.0$ (six confirmed members).
The spectroscopic members of each sub-structure occupy a circular region of comoving radius
smaller than 1\,Mpc,  which supports the existence of a physical bond among them.
This double structure is also seen in the photometric redshift analysis of a sample of 127 galaxies 
located inside a circular region of 1$^{\prime}$-radius containing the five {\it Herschel}/SPIRE sources,
where we found a double-peaked excess of galaxies at $z\simeq1.7$ and $z\simeq2.0$
with respect to the surrounding region.
These results suggest that PHz\,G95.5$-$61.6 corresponds
to two accreting nodes, not physically linked to one another,
embedded in the large scale structure of the Universe at $z\sim2$ and along the same line-of-sight.
In conclusion, the data, methods and results illustrated in this pilot project confirm that \Planck\ data
can be used to detect the emission from clustered, dusty star forming galaxies at high-$z$,
and, thus, to pierce through the early growth of cluster-scale structures.}


\authorrunning{I. Flores-Cacho et al.}
\titlerunning{Multi-wavelength characterisation of \Planck's high-$z$ dusty galaxies}
\maketitle

%INTRODUCTION
\section{Introduction}

Fully understanding the structure/galaxy formation and evolution processes
that give rise to the large scale structure (LSS) and the Universe we observe today
requires systematic studies of high-$z$ galaxies, especially at $z\sim2-3$,
where the cosmic star-formation rate (SFR) density peaks \citep{hopkins2006,sobral2013}.
We know that at these redshifts, during the formation of galaxy groups and clusters,
there is a strong link between the star formation and black hole activities.
 Feedback from these two processes affects the surrounding gas,
 which is also heated within the deepening gravitational potential wells. 
However, we do not fully understand the detailed mechanisms that made the transition
from the epoch of galaxy formation to the virialisation of today's massive dark matter (DM) halos
\citep{Mortonson2011}.
Thus, it is of crucial interest to have a complete sample of high-$z$ proto-clusters
 that  can be physically characterised.

Observing dusty galaxies at moderate to high redshifts was very demanding until it was realised that
the Rayleigh-Jeans part of the rest-frame far-infrared (FIR) spectral energy distribution (SED)
of galaxies counteracts cosmological dimming, an effect called ``negative $k$-correction''
\citep{blain1993, guiderdoni1997}.
As a consequence, the measured flux density of dusty galaxies at a fixed luminosity depends only weakly
on redshift, allowing us to detect high-$z$ (typically $2<z<6$) objects in the millimetre (mm)
and sub-millimetre (sub-mm) domains.

Bright high-redshift sub-mm and mm sources such as galaxy clusters/proto-clusters
and gravitationally lensed galaxies are relative rare on the sky.
According to \citet{Negrello2007,Negrello2010}, the surface density of  sources brighter than 300\,mJy
at 500\micron\ is $10^{-2}\,{\rm deg}^{-2}$ for strongly lensed galaxies, $3\times10^{-2}\,{\rm deg}^{-2}$
for  active galactic nuclei (AGN), and $10^{-1}\,{\rm deg}^{-2}$ for late-type galaxies
at moderate redshifts.
Other models predict similar trends \citep[\eg,][]{Paciga2009,lima2010,bethermin2011a,hezaveh2012}.
This makes even relatively shallow sub-mm surveys interesting for searches of high-$z$ objects, 
as long as they cover large parts of the sky.

Recent results from {\it Herschel} and the South Pole Telescope (SPT) demonstrated that
sub-mm and mm observations of extragalactic sources with flux densities greater than a few hundreds of mJy
at $500\,\micron$ are indeed suited to detect significant numbers of gravitationally lensed galaxies 
at high redshift \citep{Greve2010, Negrello2010,Vieira2010, Combes2012,Herranz2013, Vieira2013}.
Other predictions \citep[\eg,][]{Negrello2007} indicate that many sub-Jy sub-mm sources
could be high-$z$ clusters or groups (i.e., the progenitors of today's most massive galaxy clusters),
in line with the strong clustering of sub-mm galaxies or ultra-luminous infrared galaxies (ULIRGs)
\citep{blain2004,farrah2006,magliocchetti2007, Viero2009, Aravena2010, Amblard2011}.
Most of these objects are thought to be intensely star forming 
\citep{lefloch2005,caputi2007,pope2008,berta2011,bethermin2012}, 
in agreement with the high comoving cosmic SFR density at $z>2$ \citep{hopkins2006,Wall2008}.

In this context, the \Planck\footnote{\Planck\ ({\tiny \url{http://www.esa.int/Planck}}) is a project
of the European Space Agency (ESA) with instruments provided by two scientific consortia funded by
ESA member states (in particular the lead countries France and Italy), with contributions from NASA (USA)
and telescope reflectors provided by a collaboration between ESA and a scientific consortium led and funded by
Denmark.} satellite \citep[][]{planck2011-1.1, planck2013-p01}, with its full-sky coverage in the sub-mm and mm
over 9 frequencies can play a key role in the systematic detection of a population of high-$z$ sources.
Of particular interest here is the {\Planck} High Frequency Instrument
\citep[HFI,][]{planck2013-p03, planck2013-p03c, planck2013-p03f, planck2013-p03d, planck2013-p03e}, 
which has a broad wavelength coverage of 0.35--3\,mm with moderate resolution (between 5 and 10 arcmin).
HFI data have thus a unique potential of systematically detecting the rarest populations
of bright sub-mm galaxies across the whole sky.

Indeed, using a novel method and the full sky coverage of the \Planck\ survey,
we have identified a sample of roughly 1300 high-$z$ candidates,
that are the coldest extra-galactic sources at relatively low signal-to-noise ratio (SNR) 
close to the confusion limit \citep{Montier2014}. 
Our unique sample complements recent efforts to detect and characterise high-$z$ clusters and proto-clusters 
via their X-ray signature \citep[e.g., ][]{Gobat2011,Santos2011} or via their optical properties
\citep[typically, by detecting an over-density of red galaxies and/or tracing their red sequence\footnote{The red sequence method is limited to $z\lesssim1.5$.}, with later confirmation of spectroscopic members,
such as the works carried out by ][]{Gladders2000, Stanford2012,Brodwin2012, Muzzin2013}.
Alternatively, targeted searches around high-$z$ radio galaxies (likely to inhabit dense regions
of the LSS) have also proven to be a very powerful method of detecting high-$z$ (proto-) clusters
\citep{Venemans2005, Hatch2011a, Hatch2011b, Galametz2010b, Galametz2013,wylezalek13,rigby14}.
In this context, our mm/sub-mm identification of candidates using \Planck\ is just the first step
toward a full physical characterisation systematic follow-up observations,
in order to provide constraints on a number of actively debated open questions  in high-$z$ research,
from the dynamics of the interstellar medium (ISM) in intensely star-forming galaxies
\citep[\eg,][]{swinbank2011} to the mechanisms of structure formation \citep[\eg,][]{Mortonson2011}.

A first example of such an effort is offered by \citet{Clements2014} who,
combining the \Planck\ Early Release Compact Source Catalog \citep[ERCSC, ][]{planck2011-1.10}
and {\it Herschel} Multitiered Extragalactic Survey \citep[HerMES, ][]{Oliver2012},
selected a sample of 16 sources, four of which are confirmed structures at $0.8\le z\le 2.3$.
Given the selection process, the ERCSC is not ideally suited for high-$z$ searches
at variance from our sample of colour selected, cold sources, as shown by the high success rate
(93\%) of the dedicated follow-up of more than 200 of our candidates with {\it Herschel}/SPIRE
\citep{PlanckHerschelFU}.

This paper presents the discovery, detection and characterisation of PHz\,G95.5$-$61.6,
the first proto-cluster candidate detected in emission by \Planck\ and followed up photometrically
at optical/NIR (CFHT), MIR {\it Spitzer} and FIR/sub-mm ({\it Herschel}) wavelengths
as well as spectroscopically (VLT).
This \Planck\ source is confirmed to be associated with two galaxy systems at $z\simeq1.7$ and $z\simeq2.0$.
The paper is organised as follows. In Sect.~\ref{sec:Planck_highz}, we give a very brief description
of the \Planck\ survey as well as of our method to select proto-cluster candidates
that has led to the discovery of PHz\,G95.5$-$61.6.
In Sect.~\ref{sec:follow_up}, we present all the observations carried out to characterise PHz\,G95.5$-$61.6,
from the sub-mm to the optical, including spectroscopy.
The subsequent multi-wavelength catalogues are then described in Sect.~\ref{sec:method}.
The analysis of the available follow-up observations in terms of the over-density of colour-selected galaxies
is illustrated in Sect.~\ref{sec:analysis}, whereas the analysis in terms of photometric
and spectroscopic redshifts is presented in Sect.~\ref{sec:redshifts}.
Finally, in Sects.~\ref{sec:discussion} and \ref{sec:conclusions}, we discuss our results
and provide the conclusions of this pilot study.

% Cosmology
Throughout the paper we use a $\Lambda$CDM cosmology with
$H_0=67.3\,\mathrm{km}\,\mathrm{s}^{-1}\mathrm{Mpc^{-1}}$, $\Omega_{\mathrm{M}} = 0.315$
and $\Omega_{\Lambda} = 1 - \Omega_{\mathrm{M}}$ \citep{planck2013-p11};
all magnitudes are expressed in the AB system.

% DATA
\section{High-z sources in the \Planck\ survey}
\label{sec:Planck_highz}

\subsection{The \Planck\ survey}
\label{sec:planck}

\Planck\ \citep{tauber2010a, planck2011-1.1} is the third generation space mission
to measure the anisotropy of the cosmic microwave background (CMB).
\Planck\ observed the sky in nine frequency bands covering from 30 to 857\,GHz,
with high sensitivity and angular resolution from 31 to 5\,arcmin.
The Low Frequency Instrument \citep[LFI;][]{mandolesi2010, bersanelli2010,planck2011-1.4}
covers the 30, 44 and 70\,GHz bands with amplifiers cooled to 20\,\hbox{K}.
The High Frequency Instrument (HFI;  \citealt{Lamarre2010, planck2011-1.5})
covers the 100, 143, 217, 353, 545 and 857\,GHz bands with bolometers cooled to 0.1\,\hbox{K}.
Early astrophysics results, \linebreak based on data taken between 13~August 2009 and 7~June 2010
\citep{planck2011-1.7, planck2011-1.6}, are given in Planck Collaboration VIII--XXVI 2011.
Our results are based on the full mission data, taken between 13~August 2009 and 14~January 2012.

\begin{figure*}
\center
\includegraphics[width=\hsize]{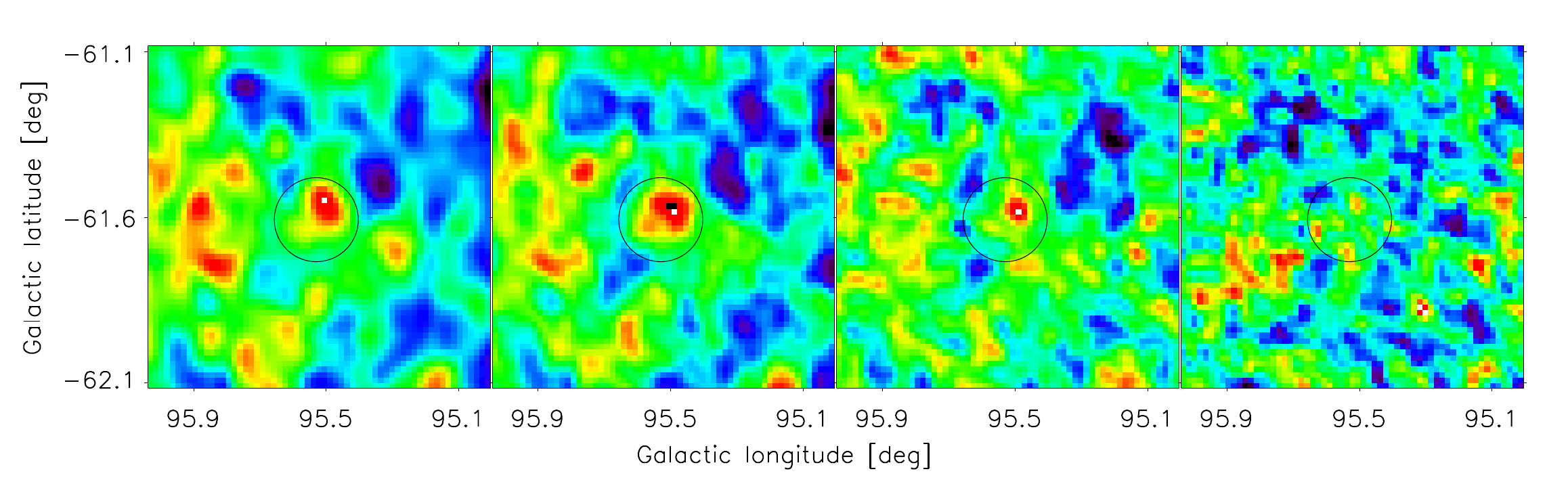} 
\caption{From left to right: {\Planck} $1\deg\times1\deg$ intensity maps at 350, 550, 850
and $1380\,\mu \mathrm{m}$  (i.e., 857, 545, 353 and 217\, GHz, respectively) centred on
the high-$z$ candidate PHz\,G95.5$-$61.6.
Their individual ranges are as follows: $\pm 0.4\, \mathrm{MJy\,sr}^{-1}$ ($350\,\mu \mathrm{m}$),
$\pm 0.2\,\rm{MJy\,sr}^{-1}$ ($550\,\micron$), $\pm 0.1\,\rm{MJy\,sr}^{-1}$ ($850\,\micron$)
and $\pm 0.05\,\rm{MJy\,sr}^{-1}$ ($1380\,\micron$).
In each panel, a circle with radius equal to the FWHM of the source (i.e., $7.2^{\prime}$)
is centred on the {\Planck} position, measured on the sky-region with excess emission
after the detection process (see Sects.~\ref{sec:detection_method} and \ref{sec:the_source}).}
\label{fig:planck_cutouts}
\end{figure*}

\subsection{Method of detection}
\label{sec:detection_method}
High-$z$ sources are extremely challenging to detect in the sub-mm,
despite their negative $k$-correction, since they appear entangled
in multiple foreground/background astrophysical emission components (such as CMB anisotropies,
the cosmic infrared background -- CIB -- fluctuations\footnote{See, e.g., \citet{planck2011-6.6}}
or Galactic emission).
Thus, it is necessary to have information in several wavelengths to extract these very cold sources
from the data.
{\it Planck}'s large wavelength coverage (from 3\,mm to $350\,\mu\mathrm{m}$)
makes it ideally suited for this task.

In our quest to identify \Planck\ high-$z$ candidates, we used the six frequency bands of {\it Planck}-HFI
and the IRIS version of the IRAS $100\,\micron$ map \citep{Miville2005},
restricting ourselves to only the cleanest (i.e., with minimal cirrus contamination
given by a hydrogen column density of $N_{\rm {HI}}<3\times10^{20}\,\rm{cm}^{-2}$) areas of the sky.
Our algorithm uses four steps, as described in full detail in \citet{Montier2014}.
First, we clean off the CMB signal in the \Planck\ maps.
We use the IRAS map as a template for the dust distribution to remove
the Galactic cirrus contamination using the {\it CoCoCoDeT} algorithm \citep[][]{montier2010}.
Then we build two excess maps (at 550 and $850\,\micron$) defined as the difference
between the \Planck\ clean map and the map resulting from interpolation
between neighbouring wavelengths assuming a power law.
Finally, we apply an optimised mexican hat wavelet detection method to identify point sources in the excess maps.

\subsection{PHz G95.5--61.6}
\label{sec:the_source}
The detection criteria previously discussed yield a sample of roughly 1300 high-$z$ candidates. 
The true nature of each catalogue entry has to be confirmed through the photometric-redshift technique
or multi-object spectroscopy.
In fact, the catalogued \Planck\ emission sources may correspond to a CIB over-density,
a single ultra-luminous or lensed galaxy, an over-density of high-$z$ dusty, star forming galaxies
or a chance alignment of lower-redshift structures along the line-of-sight (i.e., a spurious detection).
For a thorough discussion and characterisation of the \Planck\ emission sources,
we refer the interested reader to the catalogue paper \citep{Montier2014}.

PHz\,G95.5$-$61.6 (see Fig. \ref{fig:planck_cutouts}) was selected
from an earlier \Planck\ internal data release as part of a pilot programme
in order  to be followed up with {\it Herschel} due to its high SNR in the excess maps.
However, once compared against the final catalogue obtained from the full \Planck\ data,
PHz\,G95.5$-$61.6 is not an outstanding candidate, falling below the 25\% top percentile of the sample
in terms of flux density, colour and SNR
PHz\,G95.5$-$61.6 is partially resolved at high frequencies in the \Planck\ beam,
since it has a FWHM of  $7.2^{\prime}$, larger than the \Planck-HFI FWHM of $4.7^{\prime}$
at $550\,\microns$.
Its flux densities, as listed in the final catalogue \citep{Montier2014}, are: $960\pm80$,  $210\pm70$
and $1300\pm100$\,mJy at 350, 550 and $850\,\microns$, respectively.
However, we make use of the original \Planck\ maps (at $8^{\prime}$ angular resolution) here.

\section{Follow-up observations}
\label{sec:follow_up}

\subsection{{\it Herschel}/SPIRE}
\label{sec:ot2_data}

\begin{figure*}
\center
\includegraphics[width=\hsize]{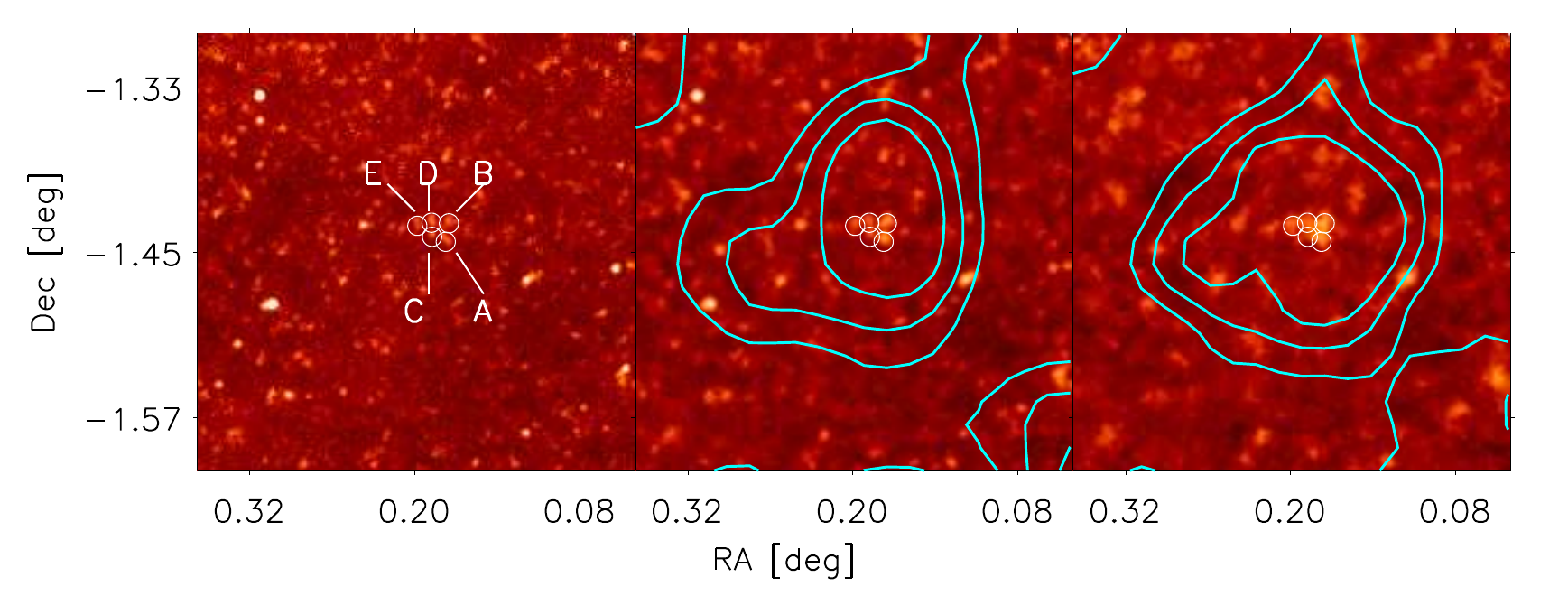} 
\includegraphics[width=\hsize]{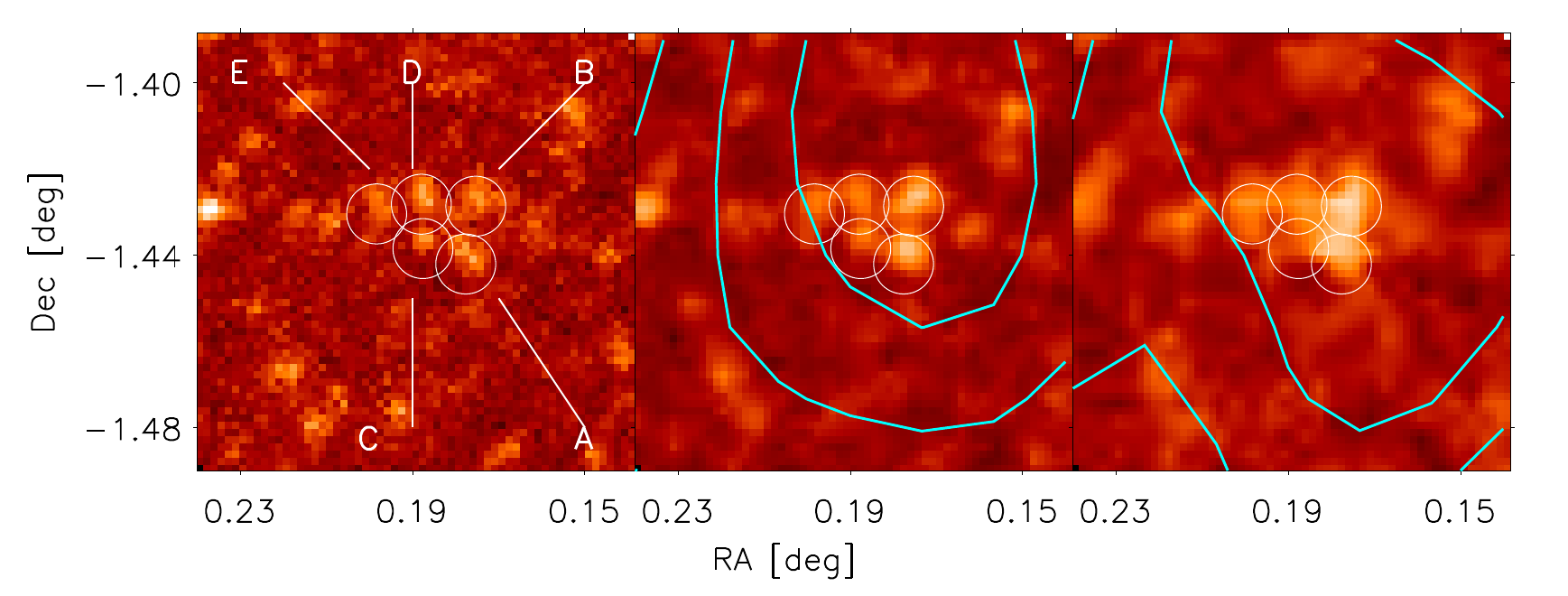} 
\caption{{\it Top row:} From left to right: {\it Herschel}/SPIRE $19\arcmin \times 19\arcmin$ images
at 250, 350 and 500\,\micron\ containing PHz\,G95.5$-$61.6.
The colour scales of these maps go from 0 to 10.0\,MJy\,sr$^{-1}$, 5.5\,MJy\,sr$^{-1}$ and 2.5\,MJy\,sr$^{-1}$ for 250, 350 and 500\,\microns, respectively.
The contours of the {\Planck} maps at 350 and 550\,\micron\ are over-plotted in cyan
onto the SPIRE 350 and 500\,\micron\ maps, respectively.
The three levels correspond to 60\%, 40\% and 20\% of the local maximum in the \Planck\ maps.
The {\it Herschel}/SPIRE point sources  within the {\Planck} $\sim5\arcmin$ beam are marked with white circles,
and labelled A, B, C, D and E on the 250\,\micron\ map.
{\it Bottom row:} A zoomed-in view to into an area of $6\arcmin \times 6\arcmin$ on source.
An additional level (80\%) has been added to the \Planck\ contours.}
\label{fig:spire_maps}
\end{figure*}

The low spatial resolution provided by the 5{\arcmin} beam of {\Planck}/HFI between 350 and 850\microns\ 
does not enable us to distinguish the individual galaxies contributing to the detected emission
at FIR/sub-mm wavelengths.
Therefore, we decided to resort to the sub-arcmin resolution offered by {\it Herschel}/SPIRE,
which enables us to confirm the {\Planck} detections, and to group our sample of high-$z$ candidates
according to their nature: either single sources (i.e., lensed galaxies candidates) or multiple sources
(i.e., galaxy proto-clusters candidates).

In 2010 we requested (PI: Montier, Programme: OT1\_lmontier\_1)
to follow up 10 of our candidates within the first open time period OT1.
Our observations were completed between 2011 and 2013,
with PHz\,G95.5$-$61.6 being the only candidate followed up in 2011.

Each SPIRE map consists of 3147\,s of observations at 250, 350 and
500\,\micron\
in the `Large Map' imaging mode.
The {\it Herschel}/SPIRE data were reduced with the Herschel Interactive Processing Environment (HIPE)
v8.0.3397, using the standard pipeline\footnote{HIPE is a joint development by the {\it Herschel} Science Ground Segment Consortium, consisting of ESA, the NASA Herschel Science Center, and the HIFI, PACS and SPIRE consortia.}.
The final SPIRE maps are the result of direct projection onto the sky and averaging of the time-ordered data
(the HIPE naive map-making routine).
Their $1\,\sigma$ noise levels are 8.1, 6.3 and $6.8\,\mathrm{mJy\,beam}^{-1}$
at 250, 350 and $500\,\mu \mathrm{m}$, respectively.
These values are close to the confusion limits, as expected by \citet{Nguyen2010}.
We rely on the {\it Herschel} calibration~\footnote{Cf. Sect. 5.2.8 of the SPIRE User Manual available at:\\ {\tiny \url{http://herschel.esac.esa.int/Docs/SPIRE/html/}}
} of the data.

The $19\arcmin \times 19\arcmin$ maps are shown in Fig.~\ref{fig:spire_maps}. 
For illustration purposes, the three SPIRE images have been resampled to match the $250\,\micron$ image.

\subsection{Spitzer/IRAC}
\label{spitzer_observations}

\begin{figure}
\center
\includegraphics[width=\hsize]{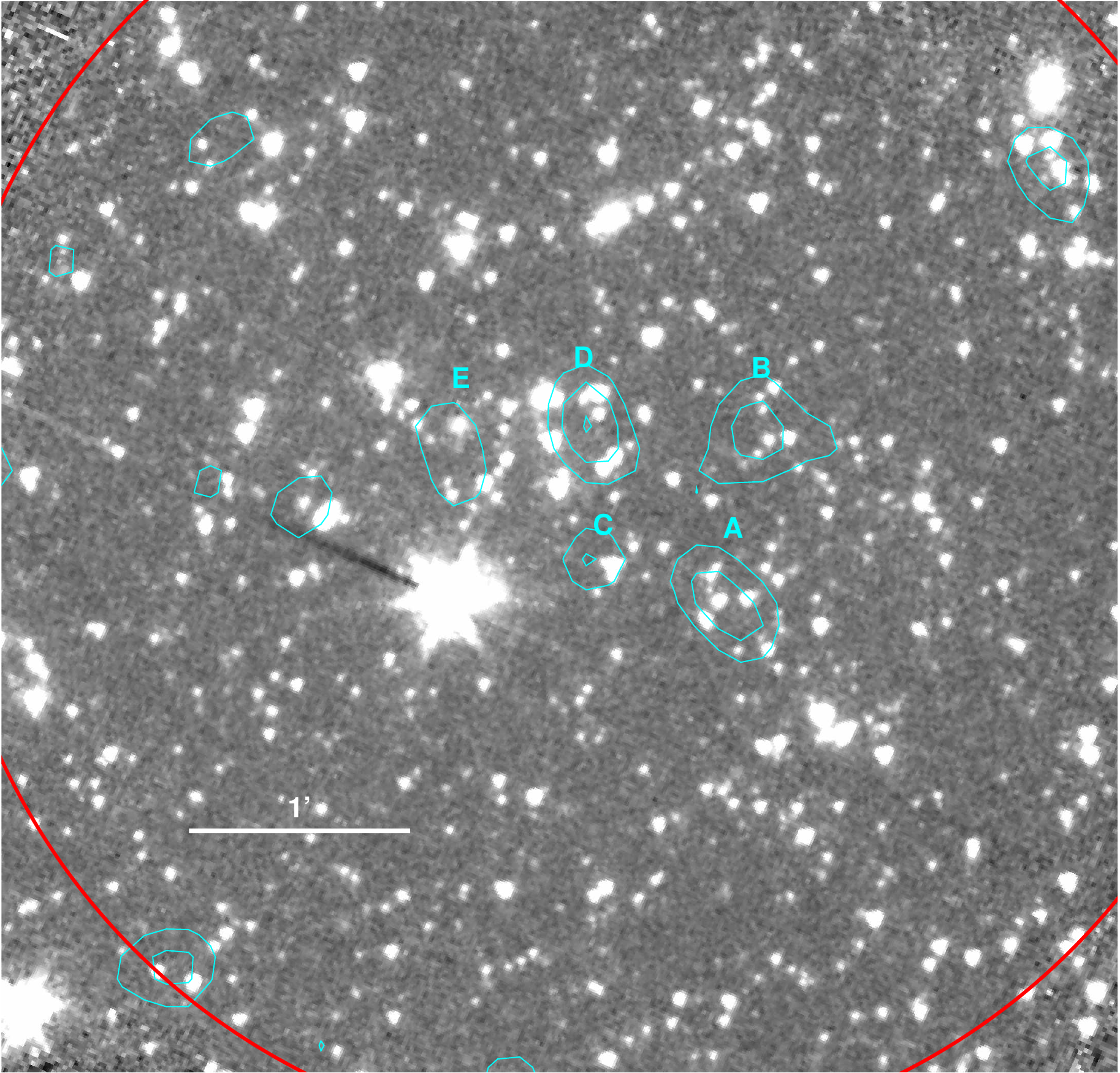}

\medskip

\includegraphics[width=\hsize]{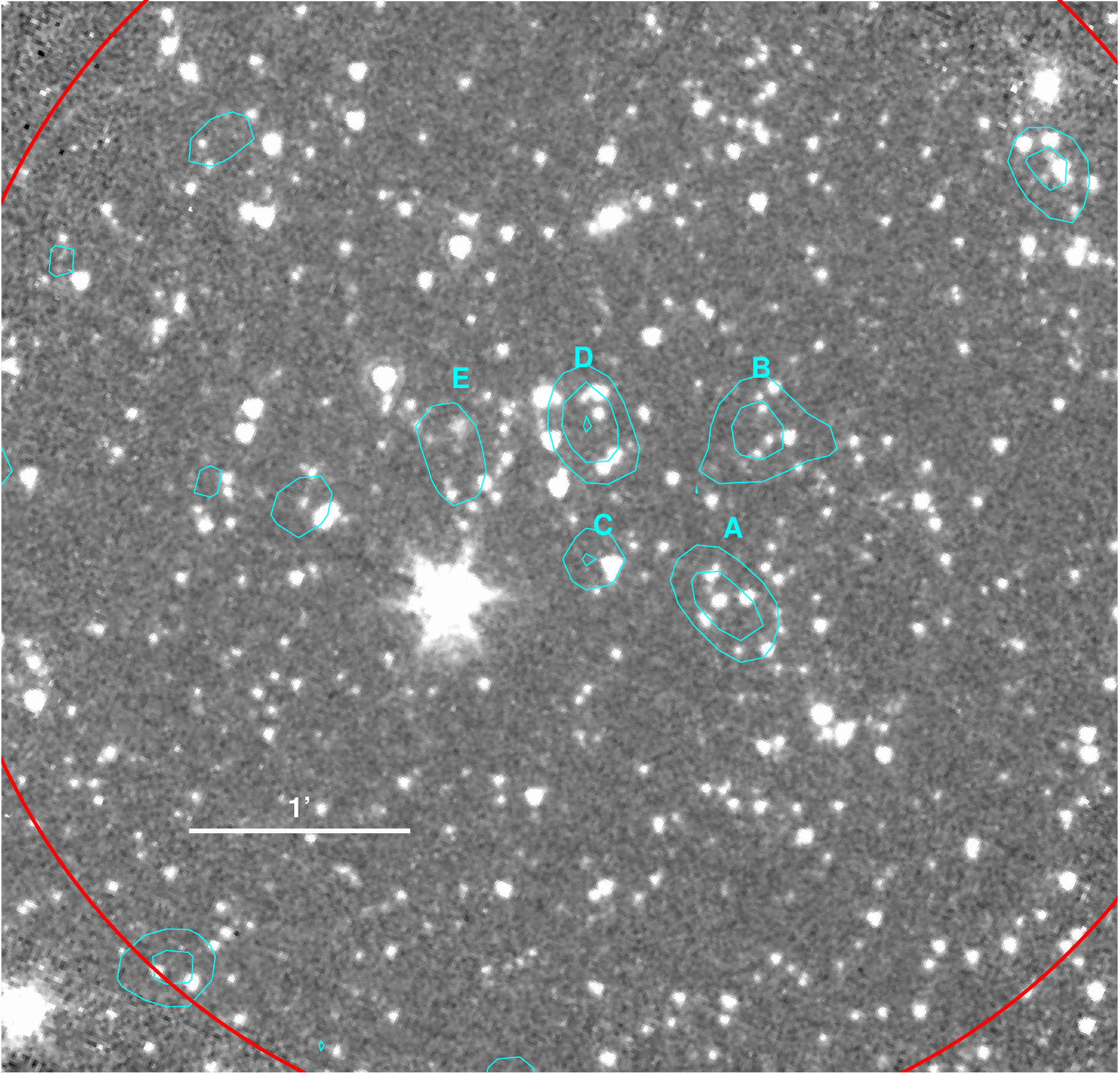}
\caption{IRAC 3.6\,\micron\ (top) and 4.5\,\micron\ (bottom) $5\arcmin\times5\arcmin$ images
around PHz\,G95.5$-$61.6, with the contours of the {\it Herschel}/SPIRE 250\micron\ map
over-plotted in cyan.
These contours correspond to deviations of 2, 3 and $4\,\mathrm{\sigma}$.
The five {\it Herschel}/SPIRE sources are labelled.
A red circle of $3^{\prime}$-radius shows the deepest area,
common to the two channels that we have used for the analysis.
North is up, East is left.}
\label{fig:IRAC_maps}
\end{figure}

We were granted 5 hours of observing time with {\it Spitzer}/IRAC \citep{Fazio2004} for broad-band imaging
($3.6\,\microns$ and $4.5\,\microns$ channels) of PHz\,G95.5$-$61.6 (plus another four high-$z$ candidates),
through programme 80238 (PI: Dole).
We observed PHz\,G95.5$-$61.6 using a 12 point dithering pattern, with 100\,s-exposure frames.
This yields a total exposure of 1200\,s per pixel at the centre of the field and 200-500\,s per pixel at the edges.
The images provided were fully reduced (level-2, post-BCDs\footnote{BCD stands for Basic Calibrated Data.})
using the IRAC Pipeline and the MOPEX \citep{Makovoz2005} software, that includes sky subtraction,
flat fielding, scattered light removal, flux calibration, correction for cosmetic defects (such as cosmic rays),
and combining the individual AORs (Astronomical Observation Requests) to a mosaic with a pixel scale
of $0.6\arcsec$.
These post-BCD have $3\,\mathrm{\sigma}$ limiting magnitudes of 23.3 and 23.1
for $3.6\,\microns$ and $4.5\,\microns$, respectively.
The final reduced images, centred at PHz\,G95.5$-$61.6, are shown in Fig.~\ref{fig:IRAC_maps}.

\subsection{CFHT/WIRCam and MegaCam}
\label{cfht_observations}

For PHz\,G95.5$-$61.6, we obtained a total of 9 hours of imaging at five optical and NIR wavelengths
with the wide-field cameras MegaCam \citep[][- $g$ and $i$ bands]{Boulade2003}
and WIRCam \citep[][- $J$, $H$ and $K_{\rm{s}}$ bands]{Puget2004}
mounted on the Canada-France Hawaii Telescope (CFHT) through programmes 11BF11 and 11BF99
(PI: Nesvadba).

All data were obtained under good and stable atmospheric conditions with seeing $<$1.0{\arcsec}
on the nights of 14~March 2012 (MegaCam) and 19~March 2012 (WIRCam).
For imaging in the $g$ and $i$ filters, the total exposure time was 1\,hour, with individual exposures of 300\,s.
Individual NIR frames had exposure times ranging from 15 to 59\,s, resulting in a total exposure of 1\,hour
for $K_{\rm s}$-band and 3\,hours for $H$- and $J$-bands.
All data were reduced using standard procedures in TERAPIX \citep[Traitement \'El\'ementaire, R\'eduction
et Analyse des PIXels,][]{Bertin2002}.

Our  final images have $3\,\mathrm{\sigma}$ limiting magnitudes (i.e., magnitude for a source
with a flux SNR of 3.0, measured within an aperture of $1.9^{\prime \prime}$-radius) 
equal to 25.7, 24.5, 24.3, 23.9, and 23.4\,AB-mag in the $g$, $i$, $J$, $H$ and $K_{\rm{s}}$ bands,
respectively.

\begin{figure}
\center
\includegraphics[width=\hsize]{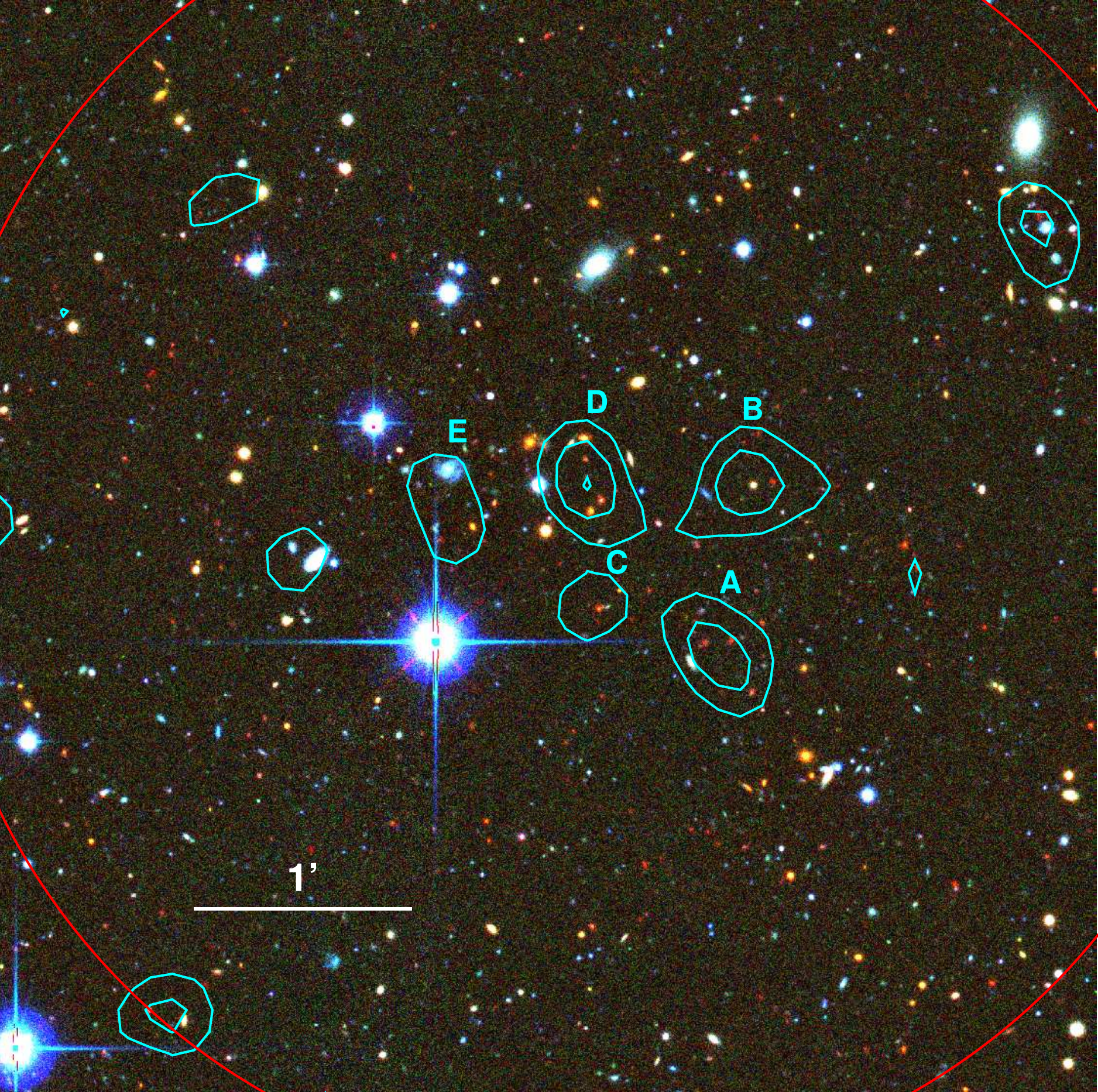} 
\caption{False colour $5\arcmin\times5\arcmin$-wide image centred at PHz\,G95.5$-$61.6,
with the RGB channels corresponding to $K_{\rm s}$, $i$ and $g$, respectively.
The contours of the {\it Herschel}/SPIRE 250\,\micron\ map are over-plotted in cyan
(see Fig.~\ref{fig:IRAC_maps}), including labels for the 5 sources in the \Planck\ beam.
North is up, East is left.}
\label{fig:false_colour_CFHT}
\end{figure}

Out of the full coverage of 1\,deg$^2$ and 0.13\,deg$^2$ from MegaCam and WIRCam, respectively,
we extracted square patches of $20\arcmin$ on each side, centred at the location
of the {\it Herschel}/SPIRE sources, which were reprojected to the MegaCam pixelation ($0.186\arcsec/$pix)
for the analysis with SWarp within the TERAPIX package.
In Fig.~\ref{fig:false_colour_CFHT}, we show a colour-composite image of the $g$, $i$ and $K_{\rm s}$ images
of PHz\,G95.5$-$61.6.

\subsection{VLT/X-Shooter}
\begin{figure}
\center
\includegraphics[width=\hsize]{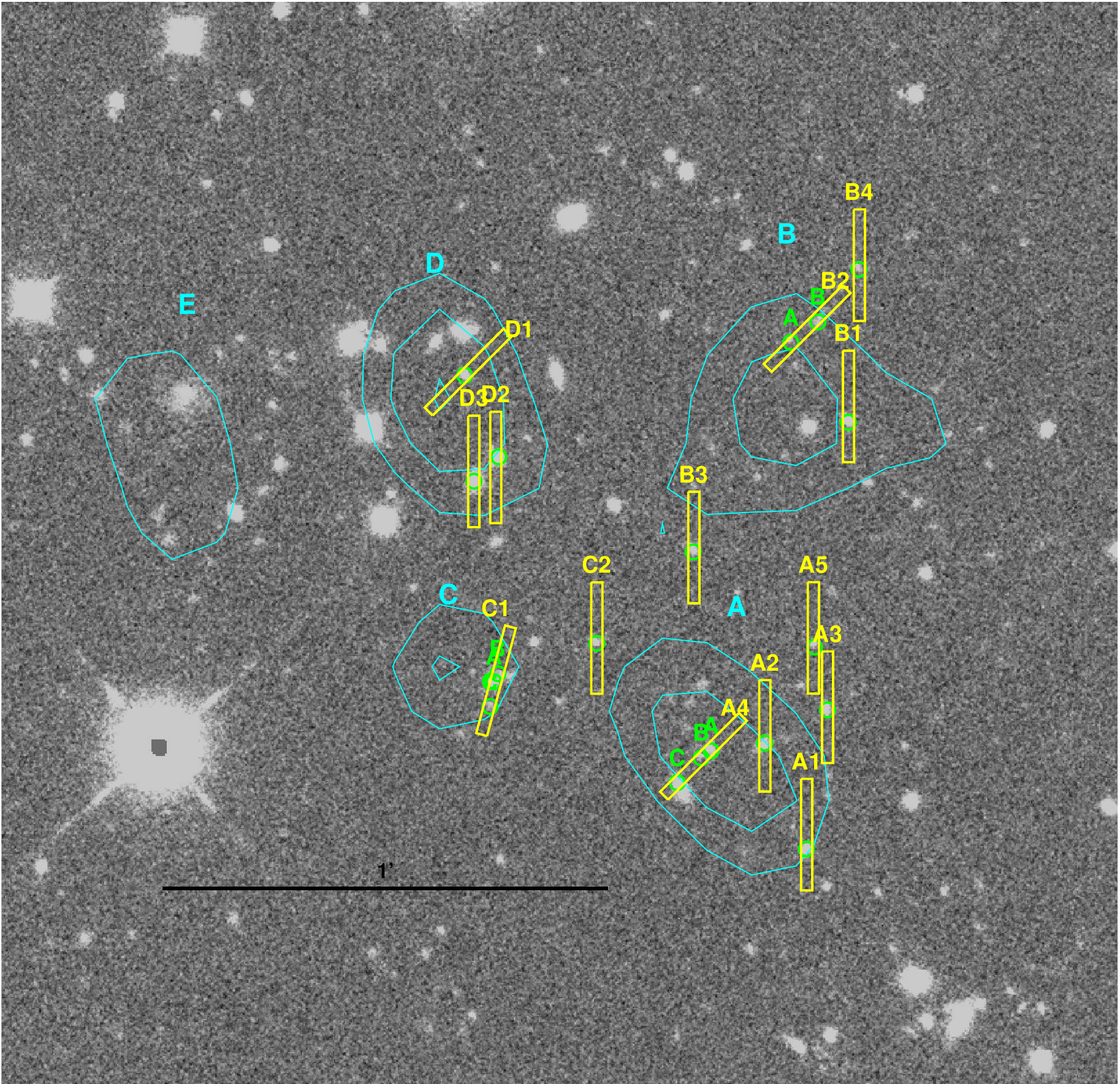} 
\caption{Positions of the slits (yellow) used for the X-Shooter observations over-plotted
onto the $K_{\rm s}$ CFHT image (grayscale).
The positions of the sources are marked with green circles, and labelled where several sources
fall in the same slit.
The contours from the SPIRE 250\microns\ image showing the FIR sources are included for reference (cyan). North is up, East is left.}
\label{fig:slits_on_ks}
\end{figure}

We have used 14~hours (programmes 287.A-5063 and 089.A-0276, PI: Nesvadba) of observing time
with the optical/NIR Echelle spectrograph X-Shooter \citep{vernet11}, mounted at the VLT,
in order to spectroscopically confirm the nature of PHz\,G95.5$-$61.6. 
We acquired long-slit spectroscopy from the ultraviolet (UV) atmospheric cutoff to the thermal cutoff
at the red end of the NIR $K_{\rm s}$ band and with a spectral resolving power $\sim$5000,
for 18 targets (see Sect.~\ref{sec:spec_analysis} for the details on the target selection).
Owing to the unique (simultaneous) spectral coverage and sensitivity of VLT/X-Shooter, various emission lines,
absorption features and/or spectral breaks were expected to be detected for galaxies deemed at redshifts
between 1.5 and 3.

Data were obtained in two runs, through programmes (PI: Nesvadba) 287.A-5063 (DDT)
and 089.A-0276.
These observations were carried out in service mode on 27-28~September 2011,
and between 4~September 2012 and 12~November 2012, respectively.
We used slits of 1.6\arcsec, 1.5\arcsec, and 1.2\arcsec\ for the UV, optical, and NIR arms, respectively,
and a $2\times1$ binning for the UV and optical arms.
For each orientation of the slit, the total exposure was equal to 50 minutes per arm
(300\,s per frame), 
and nodding mode (source at two alternating positions along the slit, separated by roughly $5^{\prime \prime}$)
was chosen to ensure a good sky subtraction in the NIR.
The full set of targets and slit orientations included in the two programs can be seen
in Fig.~\ref{fig:slits_on_ks}.
Atmospheric conditions were good and stable in both runs, with a seeing better than 1\arcsec.

We reduced the data with the ESO X-Shooter pipeline \citep[Reflex-based,
version 2.3.0][]{modigliani10, Freudling2013}, with the physical flexure modelling of \citet{bristow11}. 
One-dimensional (1-D) spectra were extracted manually, as our sources are too faint for an automatic extraction
through the pipeline.
We used a fixed aperture equal to the slit width to fold 2-D spectra into 1-D spectra.
Possible losses of flux outside the slit were not accounted for; therefore, quantities derived
from the integrated flux of spectral lines (such as the $H_{\alpha}$-based SFRs) should be considered
as lower limits.
Our manually extracted 1-D spectra were then flux calibrated (using standard stars observed each night,
their spectra being processed exactly the same way as for our sources), prior to the analysis.

For three slit orientations, we had observations from separate observing blocks (OBs),
taken several days -- or even a year -- apart).
These OBs were independently and fully reduced; the flux-calibrated 1-D spectra were then combined
into a single spectrum, using the uncertainties as weights to account for the different data quality in different days\footnote{Even if the atmospheric conditions were good and stable, there were differences in the SNR of the individual spectra.}.

% METHOD
\section{Optical and near infrared catalogues}
\label{sec:method}

\subsection{Construction of a multi-wavelength source catalogue}
\label{sec:multi_lambda_cat}

We have built a multi-wavelength catalogue of galaxies within $5^{\prime}$ from PHz\,G95.5$-$61.6
with SExtractor \citep{Bertin96} in its double image mode, using the $K_{\rm s}$ image for detection
(to favour the detection of red sources as they are more likely high-$z$ objects). 
For the optical-NIR data from CFHT, the TERAPIX reduction already provides all the images in a common grid,
which is a requirement for running SExtractor in the double image mode.
However, we had to re-project the IRAC images to the $K_{\rm s}$-image grid using SWarp
(also part of the TERAPIX pipeline).

We used the standard settings for SExtractor, with a circular aperture of $1.9^{\prime \prime}$-radius.
This large aperture is due to the moderate resolution of IRAC (FWHM$=1.7\arcsec$)
with respect to WIRCam (FWHM$=0.7\arcsec$).
Even though the IRAC handbook\footnote{See the {\href{http://irsa.ipac.caltech.edu/data/SPITZER/docs/irac/iracinstrumenthandbook/}{IRAC instrument Handbook.}}} 
establishes that the optimal aperture is $3.6\arcsec$ \citep[and this aperture has 
been used in several studies, see, e.g.,][]{Laher2012}, we deemed this aperture 
too large for our purposes, as it was prone to enclose several sources within itself.
Thus, we followed the approach of \citet{Ilbert2009} for the COSMOS data,
i.e., we adopted a fixed aperture radius of $1.9\arcsec$ for all bands.
The ensuing aperture fluxes were corrected for the loss of flux outside the aperture,
and converted into AB magnitudes using the tabulated zero-points.
Finally, all magnitudes were corrected for Galactic extinction\footnote{We obtained the average reddening
$E(B-V)$ in 2\,deg${^2}$ around the source from \url{http://irsa.ipac.caltech.edu/applications/DUST/},
using the values from \citet{Schlafly2011},
and adopted the standard visual-extinction-to-reddening ratio
$A_{v}=3.1E(B-V)$.} \citep[][]{Fitzpatrick1999}.
We added an additional flag to account for nearby sources that might contaminate the aperture flux. 

Since the portion of the sky imaged with IRAC (see Fig.~\ref{fig:IRAC_maps}) does not fully cover
the circular region with a $5^{\prime}$-radius centred on the \Planck\ position of PHz\,G95.5$-$61.6,
we restricted the analysis to a smaller region, comprised within a source-centric distance of $3^{\prime}$
(red circle in Fig.~\ref{fig:IRAC_maps}).

\subsection{Validation of the catalogue}
\label{sec:validation_cat}

In order to eliminate any possible source of bias or systematics,
we have completed a thorough validation of our multi-wavelength catalogue.
In a first step, we compared our optical/NIR data with archival data (SDSS, 2MASS),
obtaining an extremely good 1:1 correlation in all bands (the worse being J, with a median ratio
between the 2MASS photometry and ours of 0.9977), 
within the limiting magnitudes of the archival data (and selecting only non-saturated objects in our images). 
This enabled us to confirm the accuracy of the zero-points below a 0.03\,mag-error.
In addition, to guarantee that there were no systematics in zeropoints or colours,
we selected a sample of 20 isolated, non-saturated stars from the Guide Star Catalogue at CADC
(Canadian Astronomy Data Centre), and fitted their SEDs with blackbody SEDs.
The median deviations are all compatible with a null offset in zeropoint for all bands, within their individual dispersions.

As the sample of stars used to further validate the photometry was somewhat limited,
we decided to further assess the accuracy and depths of our catalogue with a MonteCarlo (MC) approach.
We injected into the final images 2000 Gaussian sources, with magnitudes ranging from 15 to 30 in all bands (fainter than the saturation level and beyond our limiting magnitudes), with a point-spread-function (PSF)
matching the experimental FWHM of the images.
The coordinates for the injection of these {\it mock} sources were randomly selected across a $20\arcmin$ patch 
around PHz\,G95.5$-$61.6.
Then we ran SExtractor and compared recovered magnitudes with input magnitudes.
This process was repeated 3 times.

Image depths (i.e., limiting magnitude) were confirmed with this MC simulation,
obtaining recovery rates of 99\% for magnitudes of 25.0, 23.9, 23.6, 23.4 and 22.8\,AB-mag
in the $g$, $i$, $J$, $H$ and $K_{\rm{s}}$ bands, respectively.
These are slightly fainter (0.1 to 0.2\,mag) than our estimates based on flux,
most likely due to the loss of sources falling on top of (or near to) pre-existing sources in the images. 

\begin{figure}
\center
\includegraphics[width=\hsize]{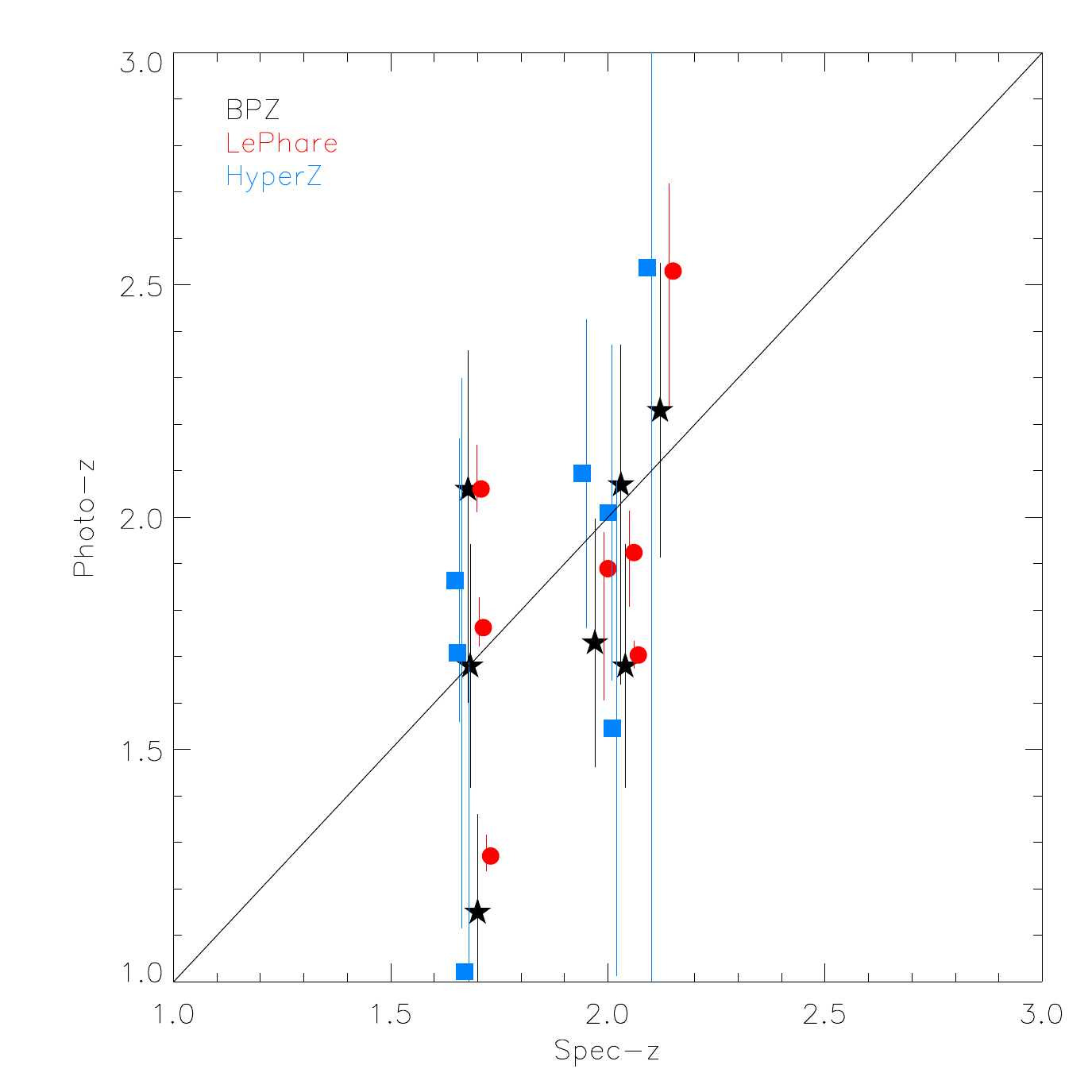} 
\caption{Comparison of the photometric redshift estimates derived with the BPZ method of template fitting against the spectroscopic redshift obtained from X-Shooter observations. For consistency check we also used two other photo-$z$ codes, namely HyperZ \citep{Bolzonella2000} and {\it LePhare} \citep{Arnouts1999, Ilbert2006}), which results are over-plotted.  Note that they have been shifted horizontally by $\pm0.03$ to ease the visualisation. These two codes show similar results with slightly higher dispersion so they are not used in the rest of the paper. }
\label{fig:photoz_comparison}
\end{figure}

\subsection{Photometric redshift calibration}
\label{sec:calibration_photoz}

Photometric redshifts are the right tool to assess the nature of the structure(s) causing the signal
detected by \Planck . 
We used the BPZ code \citep{Benitez2000} of photo-$z$ estimation. 
It is based on template fitting, using a multi-wavelength catalogue. 
In order to estimate the accuracy of the resulting photo-$z$'s, we selected seven of the 13 galaxies
for which we obtained reliable redshift measurements with X-Shooter.
These seven sources have very clean photometry (neither contamination from nearby sources
nor flags in the SExtractor photometry). However, our photometric redshift estimations show large uncertainties, due to broad and/or multi-peaked probability distributions in the photo-$z$ space (Fig. \ref{fig:photoz_comparison}). The mean dispersion (given by $\langle|z_{\rm spec}-z_{\rm photo}|/(1+z_{\rm spec})\rangle$) is $\sim8.5\%$, a value fully acceptable in this high redshift range, although we are fully conscious of the limitations of method even in this best-case scenario. 

% ANALYSIS
\section{Galaxy over-density in colour}
\label{sec:analysis}
\subsection{Red excess in the sub-mm}

\begin{figure}
\center
\includegraphics[width=\hsize]{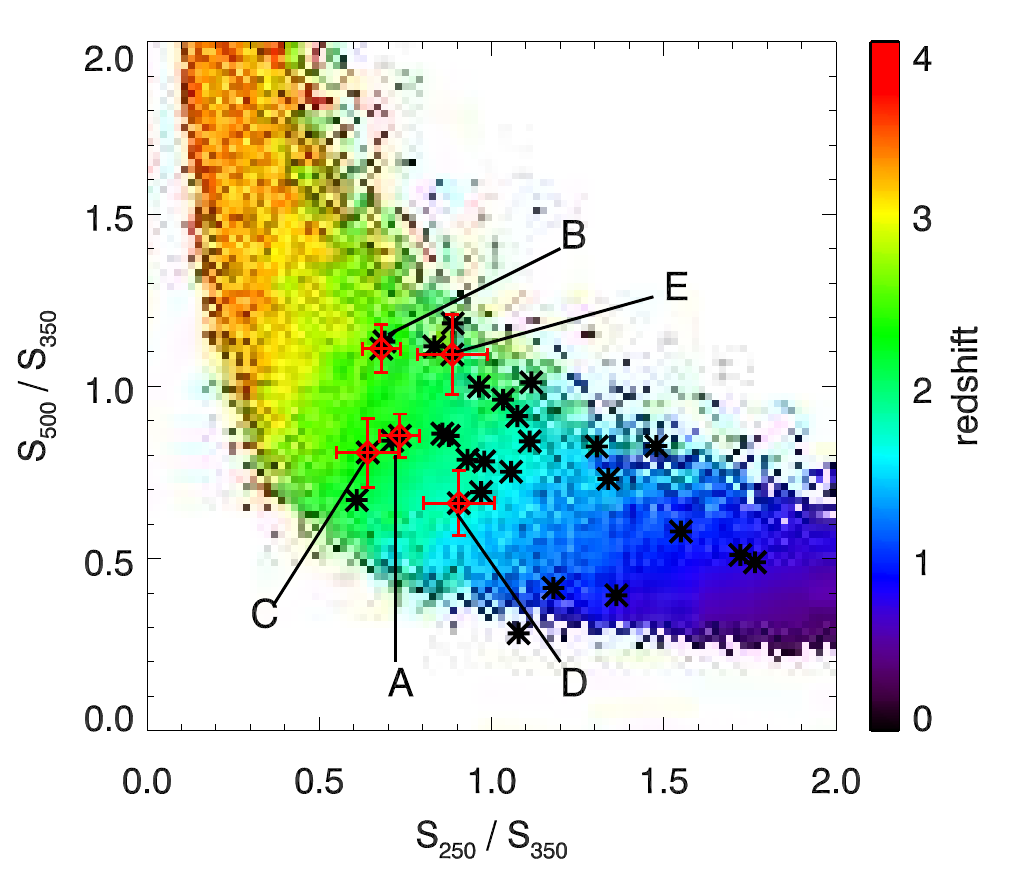} 
\caption{Colour-colour diagram of {\it Herschel}/SPIRE sources
in the region of PHz\,G95.5$-$61.6.
The coloured background indicates the average redshift in this colour-colour space
for $10^6$ randomly generated synthetic SEDs \citep{Amblard2010}, modelled as modified blackbodies
with dust spectral indices ranging from 1 to 2, and dust temperatures ranging from 10\,K to 40\,K. 
Sources detected above $2.5\,\mathrm{\sigma}$ in all SPIRE bands are plotted as black stars.
The five {\it Herschel}/SPIRE sources at $250\,\micron$ located around PHz\,G95.5$-$61.6,
within the \Planck\ beam, are marked with red diamonds and labelled A, B, C, D and E,
as in Fig.~\ref{fig:spire_maps}.}
\label{fig:cc_spire}
\end{figure}

The {\Planck} data for PHz\,G95.5$-$61.6 (see Fig.~\ref{fig:planck_cutouts}) reveal a clear excess
in surface brightness at $550\,\micron$, typical of a high-$z$ source, as dust emission normally peaks
at $100-160\,\micron$ (rest-frame).
From the {\it Herschel} images, we have extracted a catalogue with the FASTPHOT software
\citep{Bethermin2010}.
Imposing a SNR threshold of 2.5 in all SPIRE bands, our catalogue comprises 28 sources
within 12{\arcmin} from the image centre,
out of which five are within the \Planck\ beam and are labelled A,
B, C, D and E in the left panel of Fig.~\ref{fig:spire_maps}.
This large number of SPIRE sources clustered around the position of PHz\,G95.5$-$61.6,
within the Planck beam, represents a deviation of $3.4\,\mathrm{\sigma}$ with respect to 
the rest of the field in terms of the number density of SPIRE sources.
  
The colour-colour plot of the {\it Herschel}/SPIRE sources in the region of PHz\,G95.5$-$61.6
(see Fig.~\ref{fig:cc_spire}) was compared with the loci of average redshift obtained by \cite{Amblard2010}
for $10^6$ randomly generated SEDs.
Following the prescription of \citet{Amblard2010}, these simulated SEDs correspond
to modified blackbody SEDs with a spectral index ranging from 1 to 2, and dust temperature
ranging from 10\,K to 40\,K.
The five sources located within the \Planck\ beam (marked with red diamonds in Fig.~\ref{fig:cc_spire}) exhibit colour-colour ratios consistent with those of $z\sim2$ sources. This overdensity could in fact be the densest part of an even larger structure \citep{muldrew15}.

\begin{figure}
\center
\includegraphics[width=\hsize]{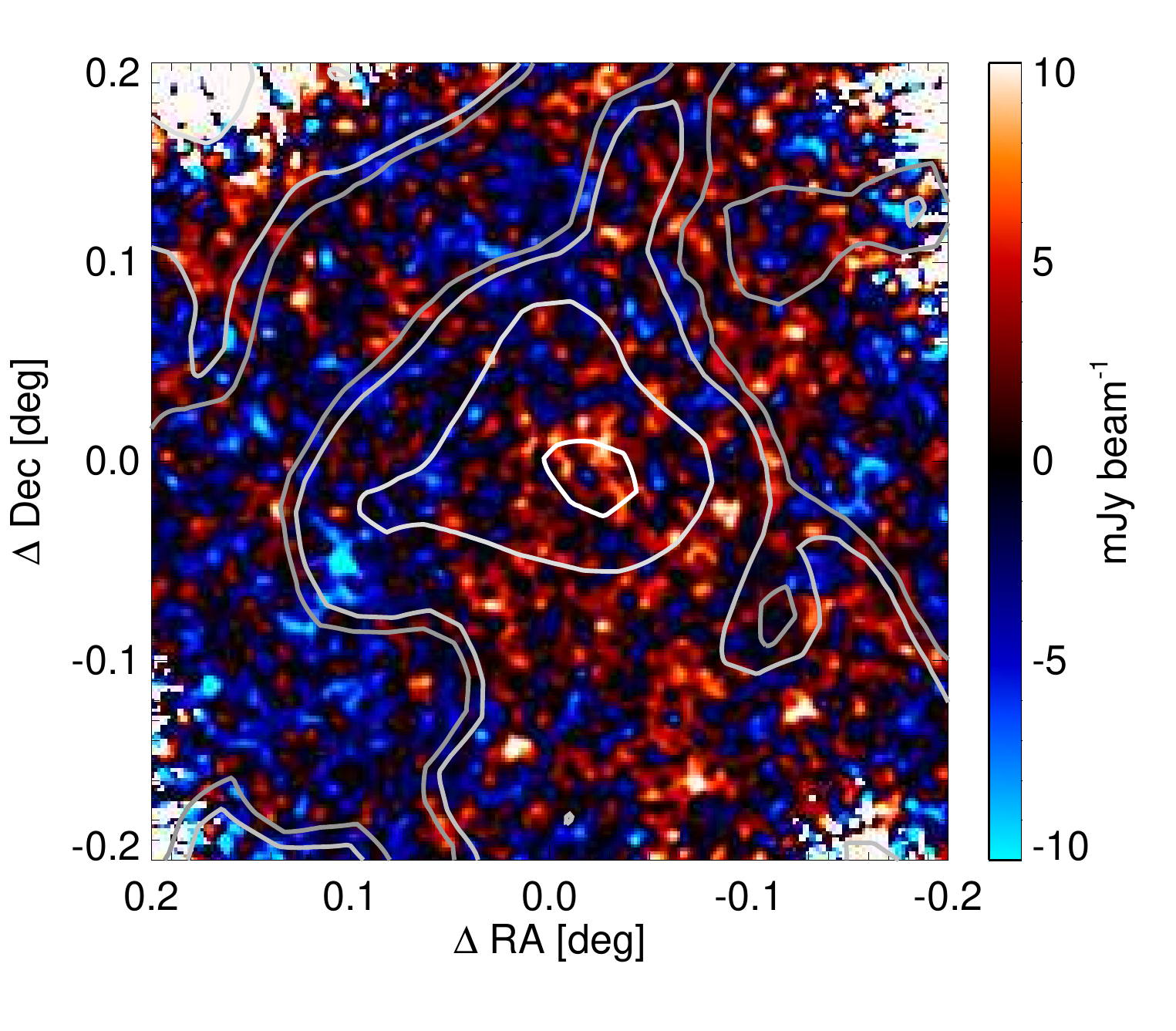} 
\caption{SPIRE ``red excess'' map ($M_{\rm RX}$) at $500\,\micron$. The
  contours of the cleaned {\Planck} map at $550\,\micron$ are overlaid  as
  grey lines at four levels: 0, 10, 50 and 90\% of the local
  maximum.}
\label{fig:red_excess_map}
\end{figure}

We have constructed a ``red excess'' map ($M_{\rm RX}$)
from the {\it Herschel}/SPIRE $350$ and $500\microns$ maps ($M_{350}$ and $M_{500}$,
respectively) defined as:
\begin{equation}
M_{\rm{RX}} = M_{500} - M_{350} \times \left< \frac{M_{500}}{M_{350}} \right>_{\rm bkg},
\end{equation}

\noindent where $ \left< M_{500}/M_{350} \right>_{\rm bkg}$ is the average colour
computed across the background area (defined as the region beyond the FWHM of the source flux profile
measured in the \Planck\ 550\microns\ map).

Positive fluctuations of $M_{\rm{RX}}$ trace structures with redder colours than the average value of the CIB,
which can be explained by higher redshifts or colder SEDs.
These ``red'' fluctuations may correspond to single galaxies at high redshift,
clumps of coeval galaxies or alignments of faint galaxies distributed at various redshifts
but along the same line-of-sight.

The SPIRE ``red excess'' map at $500\,\micron$ is shown in Fig. \ref{fig:red_excess_map}.
We find a strong ``red excess'' at the position of the \Planck\ source (shown as a contour map):
an over-density of ``red'' structures in the {\it Herschel}/SPIRE maps
is spatially correlated with PHz\,G95.5$-$61.6.
However,  the red excess map shows other peripheral red peaks that are not associated
with our \Planck\ source.
We have assessed the significance of the central ``red'' peak against MC simulations
performed on the public HerMES fields\footnote{All public HerMes data are available via the HeDaM website: {\url {http://hedam.oamp.fr}}.} \citep{Hermes2012}.
The central ``red excess'' is significant at $\sim 2.7\,\mathrm{\sigma}$.
For details on the computation of this significance see \citet{Montier2014}.

\begin{figure}
\center
\includegraphics[width=\hsize]{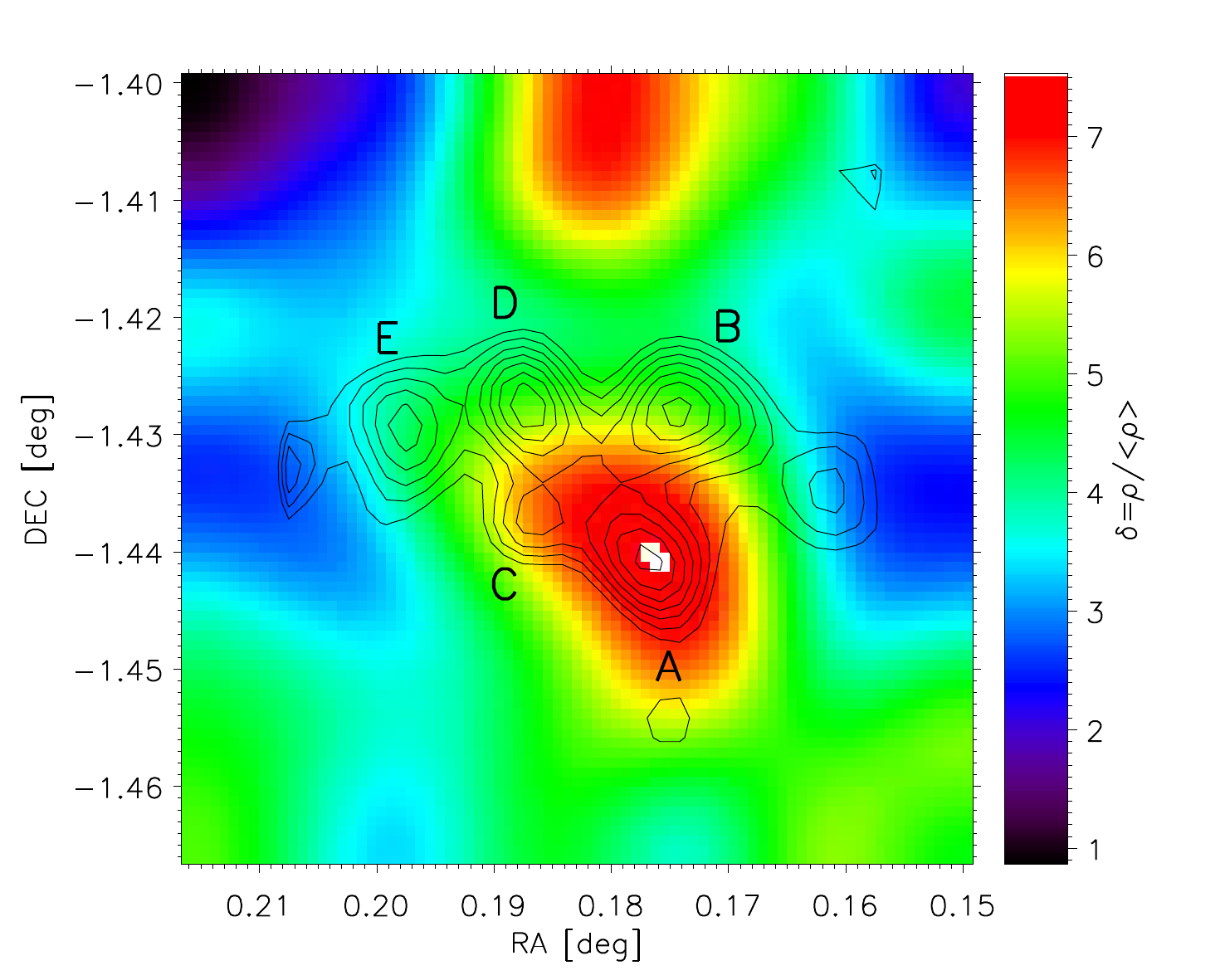}
\caption{Over-density maps of galaxies with colours 
$1.75<i-K_{\rm s}<3.25$. The {\it Herschel}/SPIRE contours are over-plotted in black, 
  showing the location of the SPIRE sources (labelled).}
\label{fig:cfht_overdensity}
\end{figure}

\subsection{Optical and NIR over-density of red galaxies}
We used the multi-wavelength photometric catalogue of galaxies built out of the CFHT and {\it Spitzer} imaging
of PHz\,G95.5$-$61.6 to compute over-density maps of colour-selected galaxies at 1{\arcmin}-resolution.
The over-density map for a given colour range is given by the ratio
between the galaxy number density map and the average galaxy number density computed
for this colour interval in the field, i.e., across a square area of 15{\arcmin} on each side,
masking a square area of 4{\arcmin} on each side centred on PHz\,G95.5$-$61.6. Each number density map is computed using the 10th nearest neighbour method. 
It reveals an over-density of (likely) high-$z$ galaxies centred on the {\it Herschel}/SPIRE source A, which exhibits an emission peak 1.75 times higher
than the average emission in the field,
and represents a robust deviation ($2.9\,\mathrm{\sigma}$) from the field.

\begin{figure}
\center
\includegraphics[width=\hsize]{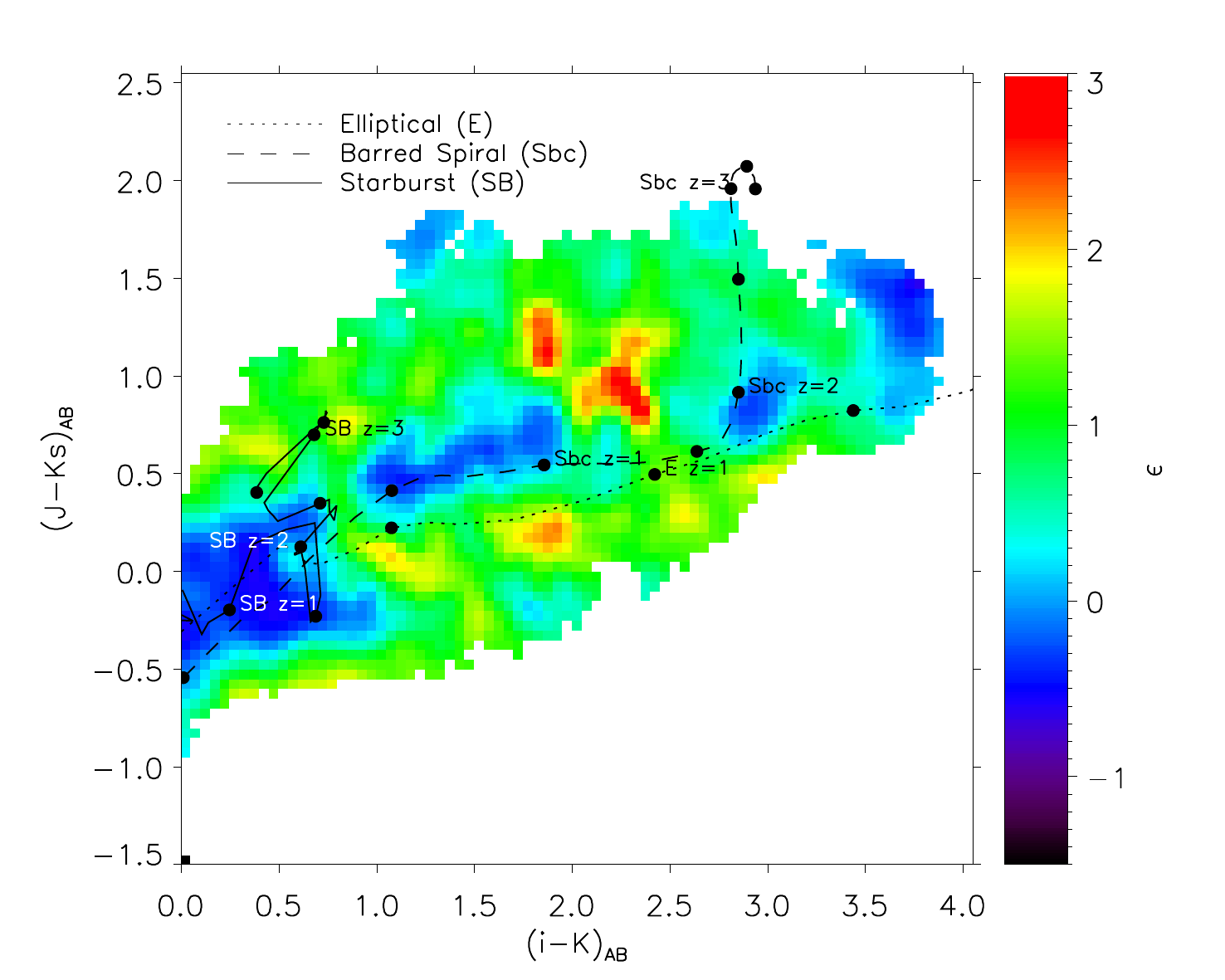} 
\caption{Comparison of the galaxy distribution in the $i-K_{\rm s}$ versus $J-K_{\rm s}$ diagram
for the circular region of $1\arcmin$-radius containing the five {\it Herschel}/SPIRE sources
and a region external to it (i.e., the `field').
An excess ($\epsilon$, see Eq.~\ref{eq:delta}) of galaxies with $i-K_{\rm s}\sim2.3$
and $J-K_{\rm s}\sim0.8$ is detected at the location of PHz\,G95.5$-$61.6 with respect to the field
(see text for further details). For visualisation purposes, the map has been smoothed with a gaussian beam of two pixels.
Tracks describing the colours for three typical SED templates of local galaxies, observed from $z=0$ to $z=4$, are over-plotted.
For each track, filled black circles mark the loci at a redshift step of 0.5, with added labels at redshifts 1, 2 and 3.}
\label{fig:colourcolourplane}
\end{figure}

When using a selection based on $i-K_{\rm s}$, we clearly detect a strong over-density of galaxies 
with $i-K_{\rm s}>1.8$\footnote{We note that $i-K_{\rm s}>2.61$ defines Extremely Red Objects at $1\le z \le2$ \citep{Pozzetti2000}} centred on the {\it Herschel}/SPIRE source A (Fig.~\ref{fig:cfht_overdensity}). The average density of such red galaxies is roughly 7 times higher than the field one
($10\,\mathrm{\sigma}$ deviation) near Herschel source A.

In addition, the distribution of galaxies in the ($J-K_{\rm s}, i-K_{\rm s}$) colour-colour diagram
suggests the presence of an over-density of dusty starbursts at $1 \le z \le 2$ \citep{Pozzetti2000, Pierini2004}.
To analyse the over-density of red sources around the position of PHz\,G95.5$-$61.6
in this colour-colour diagram, we have divided the colour-colour space in a regular grid
with a bin of 0.05\,mag and computed the number density at each pixel
via the nearest-neighbour method.
We have applied this method to two separate samples based on their distance
to the {\it Herschel}/SPIRE sources: i) an `on-source' sample that includes all sources
located within $1.0\arcmin$ of PHz\,G95.5$-$61.6 and yields number density $\rho_{\rm source}$;
ii) a control (or `field') sample with objects within an annulus of inner and outer radii
equal to $1.5\arcmin$ and $4.0\arcmin$, respectively, that yields a number density $\rho_{\rm field}$.
The excess map  is then given by:
\begin{equation}
  \label{eq:delta}
  \epsilon = \dfrac{\rho_{\rm source} - \rho_{\rm field}}{\rho_{\rm field}}, 
\end{equation}

In Fig.~\ref{fig:colourcolourplane}, there is a strong excess (by 3.8 times) of colour-selected sources 
($i-K_{\rm s}\sim2.35$, $J-K_{\rm s}\sim0.80$) spatially correlated with PHz\,G95.5$-$61.6. %,
For reference, we compare this map to the tracks describing the colours obtained
for three template SEDs of local galaxies observed at a redshift from 0 to 4.
The elliptical galaxy template and the Sbc spiral galaxy template come from \citet{Coleman1980},
whereas the starburst galaxy template comes from the \citet{Bruzual1993} library
as described in \citet{Benitez2004}.
Even without any correction for dust attenuation to these redshifted tracks, 
the position in the colour-colour diagram of the detected excess is consistent with
the presence of high-$z$ galaxies. It also suggests that the excess is not dominated by starburst galaxies and that most of the galaxies are more standard objects, with some possible dust redenning.

\section{Redshift measurements}
\label{sec:redshifts}
\subsection{Photometric redshift results}
\label{sec:photoz}

We applied BPZ \citep{Benitez2000} to our photometric catalogue
in order to obtain photometric redshifts of all sources.
A library of 30 SED templates was used to reproduce the observed multi-wavelength photometry.
We did not impose any prior on the data.
Together with an estimation of the photometric redshift ($z_{\rm photo}$),
BPZ yields the redshift probability distribution function for each source,
as well as a two-fold assessment of the reliability of this estimation
via the $\chi^2$ of the fit and the odds that the solution of the fit is unique.

\begin{figure}
\center
\includegraphics[width=\hsize]{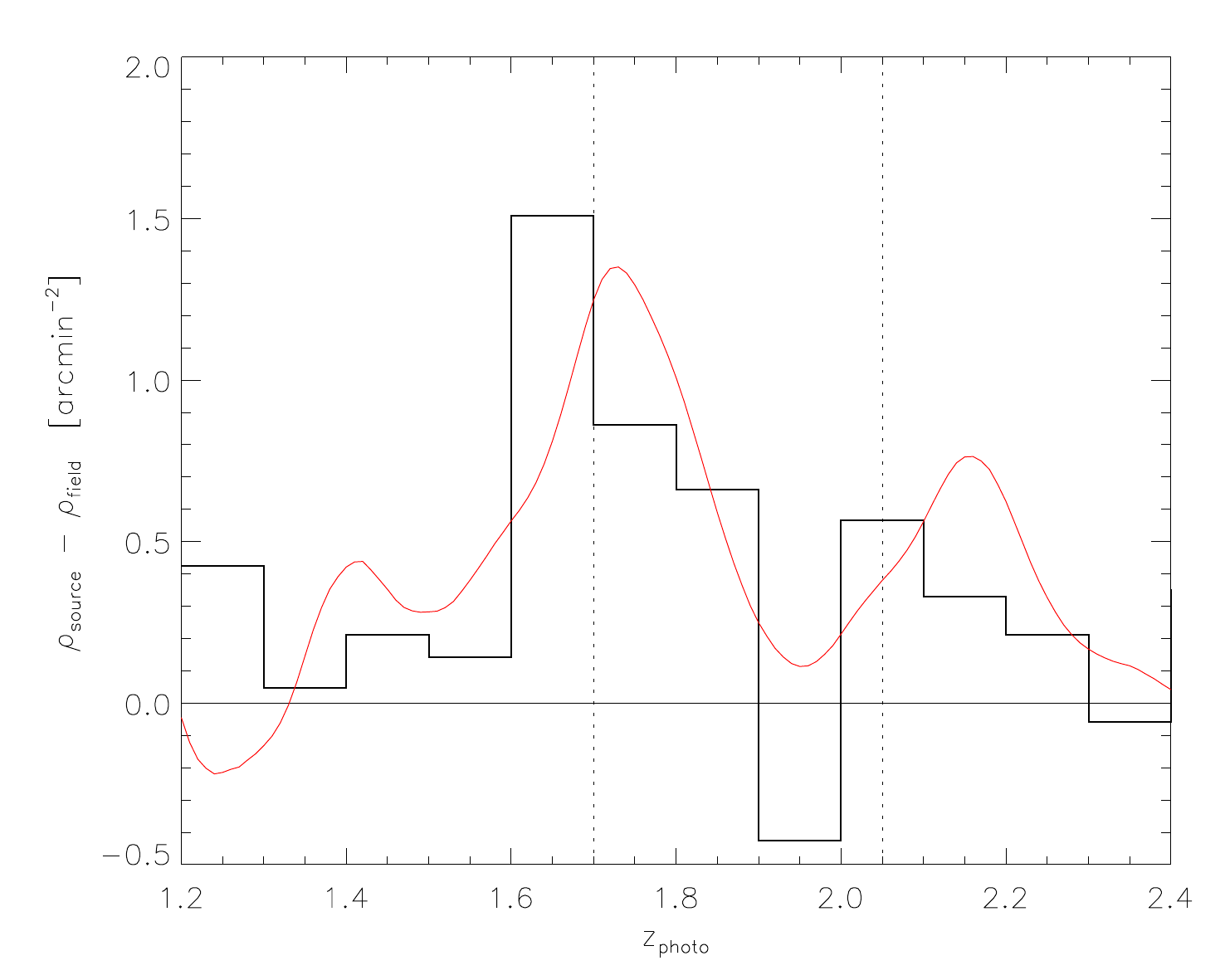} 
\caption{Excess in the photometric redshift distribution of galaxies within $1\arcmin$
from PHz\,G95.5$-$61.6 with respect to that obtained for the rest of the field. The red line is the sum of the photometric redshift probability distribution functions for the same objects.
The existence of a broad peak, consistent with the two systemic redshifts spectroscopically obtained
from the X-Shooter data (vertical dashed lines) is suggested, with about 20 sources in this overdensity.
}
\label{fig:histo_photoz}
\end{figure}

We computed the galaxy distribution in $z_{\rm photo}$ (with a bin size of $\Delta z_{\rm photo}=0.1$)
for two regions of the sky: a circular region of $1\arcmin$-radius centred on source
and its adjacent circular annulus with outer radius of $3\arcmin$.
After rescaling to the same area, we find an excess of sources at high-$z$ around PHz\,G95.5$-$61.6  between $z_{\rm photo}\simeq1.60$ and $z_{\rm photo}\simeq2.20$ (Fig.~\ref{fig:histo_photoz}), 
in good agreement with the results obtained from the galaxies observed with X-Shooter
(see Sect.~\ref{sec:spec_analysis} below).
However, given the large uncertainties
in the photo-$z$ estimations, we cannot, from these data alone, establish whether the excess 
of high-$z$ galaxies is caused by two separate systems or a broad one.  

\subsection{Spectroscopic confirmation of the high-z nature of PHz G95.5--61.6}
\label{sec:spec_analysis}

We obtained optical/NIR spectra of 13 targets (shown in Fig.~\ref{fig:slits_on_ks},
which were chosen using three general criteria:
\begin{itemize}
\item radial distance to one of the {\it Herschel}/SPIRE $250\,\micron$ sources $< 25\arcsec$;
\item $K_{\rm s} > 19 $;
\item $J-K_{\rm s} \ga 1$).
\end{itemize}
Specifically, these colour criteria favour the selection of galaxies with $1.0<z_{\rm photo}<3.4$
when applied to the COSMOS photometric redshift
catalogue\footnote{\url{http://irsa.ipac.caltech.edu/cgi-bin/Gator/nph-dd}} \citep{Ilbert2009},
consistent with our target galaxies.
There are two exceptions to the colour requirement: B2b and A4c, which were simply targets of opportunity, observed by tilting the slit.
In addition, C1a and C1b are fully blended in our aperture photometry.

\begin{figure}
\center
\includegraphics[width=\hsize]{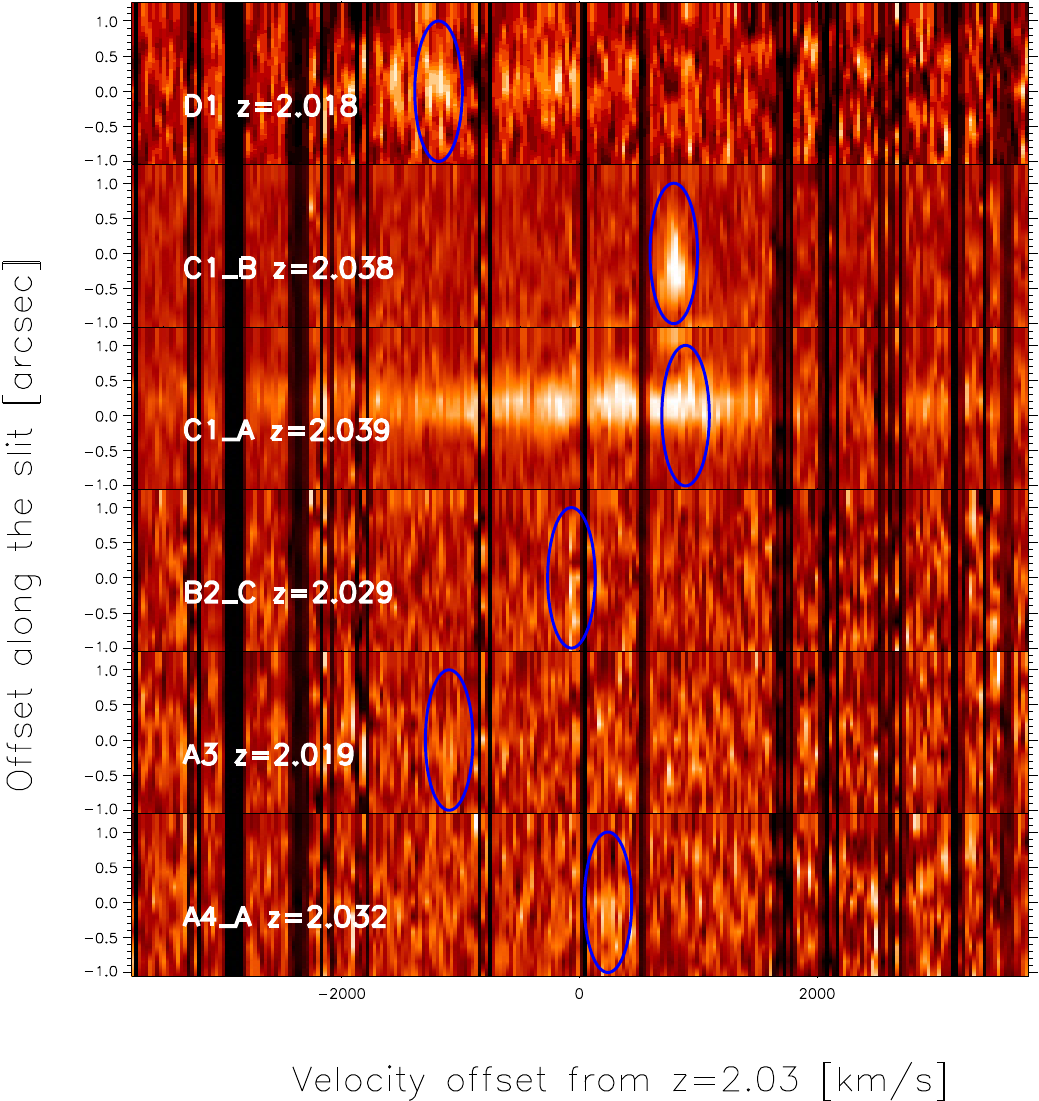}
\includegraphics[width=\hsize]{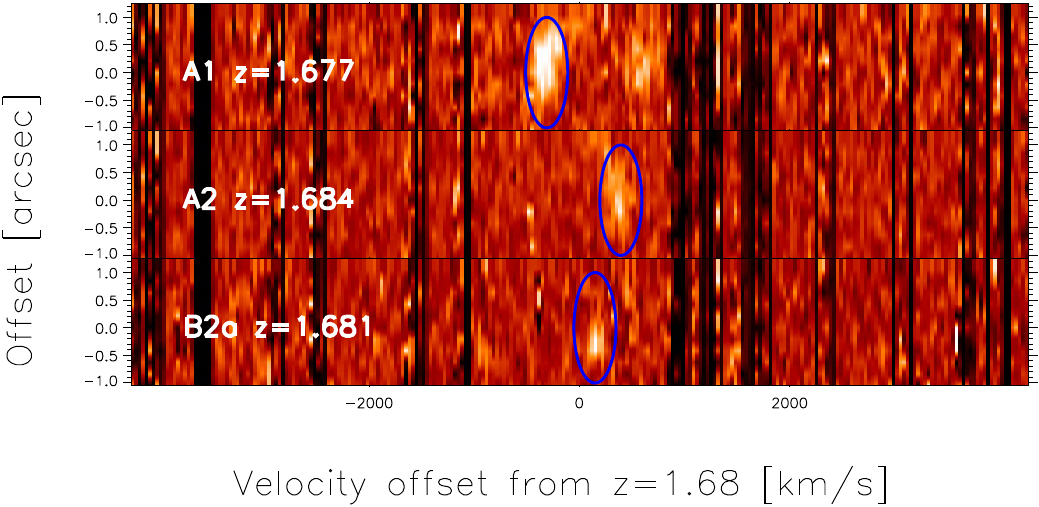} 
\caption{Zoomed-in view to the 2-D spectra obtained with VLT/X-Shooter
for those galaxies that are spectroscopically identified as members of any of the two galaxy systems
set at (systemic redshift) $z\simeq2.03$ (top panel) and $z\simeq1.68$ (bottom panel).
The spectra are shifted accordingly to show the velocity offset with respect to the systemic recession velocity
in correspondence of the $\rm{H}_{\alpha}\lambda6564$ emission line,
marked with a blue ellipse.
For A1, and D1, the contiguous [NII]$\lambda$6584 
emission line can also be (hardly) seen. For C1a the $\rm{H}_{\alpha}$ emission line is broad and  blended with some contribution from the [NII]$\lambda$6550 and $\lambda$6584 lines, for a total width of more than 200\AA . 
For display purposes, known sky lines have been masked, 
with a width of 0.2\,nm, and the colour range has been adapted
for each source to maximise clarity.
However, we have consistently identified the spectral features in non-masked versions of the spectra.}
\label{fig:twodspec}
\end{figure}

Our observations provided reliable redshift estimations for 13 targets 
(listed in Table \ref{tab:mags_17p732}, with magnitudes in the optical, NIR and MIR bands), 
all with redshifts between $\sim1.61$ and $\sim2.1$.
Most of these redshift determinations arise from two or more spectral features.
The exceptions are A3, A4a, A4c, and B2b, for which we reasonably assume that
the brightest (single) emission line that we see is $H_{\alpha}$.

\begin{table*}
\centering
\caption{Positions, magnitudes at optical (MegaCam), NIR (WIRCam) and MIR (IRAC)
wavelengths, and spectroscopic redshifts of the 13 galaxies followed-up with VLT/X-Shooter. All magnitudes are expressed in the AB system. For clarity, the table contains three groups: members of the structure at $z\simeq1.7$, members of the structure at $z\simeq2.0$, and non-members of either group (and, in all cases, ordered from lower to higher redshift). 
}
\label{tab:mags_17p732}
\nointerlineskip
\setbox\tablebox=\vbox{
\newdimen\digitwidth % These five lines change what an asterisk
\setbox0=\hbox{\rm 0} % means to TeX. Instead of meaning
\digitwidth=\wd0 % "print an '*' here", it now means "leave
\catcode`*=\active % as much blank space as a single number
\def*{\kern\digitwidth} % takes up".
\newdimen\signwidth % These five lines change the meaning of an
\setbox0=\hbox{{\rm +}} % exclamation mark in the same way, so that it
\signwidth=\wd0 % leaves as much space as a plus or minus sign.
\catcode`!=\active % These definitions will disappear at the end of
\def!{\kern\signwidth} % the \vbox.
\newdimen\pointwidth % These five lines change the meaning of a
\setbox0=\hbox{.} % question mark in the same way, so that it
\pointwidth=\wd0 % leaves as much space as a decimal point.
\catcode`?=\active % These definitions will disappear at the end of
\def?{\kern\pointwidth} % the \vbox.
\halign{
#\hfil\tabskip=1.0em&
\hfil#\hfil\tabskip=1.0em&
\hfil#\hfil\tabskip=1.0em&
\hfil#\hfil\tabskip=1.0em&
\hfil#\hfil\tabskip=1.0em&
\hfil#\hfil\tabskip=1.0em&
\hfil#\hfil\tabskip=1.0em&
\hfil#\hfil\tabskip=1.0em&
\hfil#\hfil\tabskip=1.0em&
\hfil#\hfil\tabskip=1.0em&
#\hfil\tabskip=1.0em&
\hfil#\hfil\tabskip=0pt\cr
\noalign{\doubleline}
ID&RA&Dec&$g$&$i$&$J$&$H$&$K_{\rm s}$&IRAC&IRAC&$z_{\rm spec}$\cr
 & & & & & & & & $3.6\microns$ & $4.5\micron$ & \cr
\noalign{\vskip 5pt\hrule\vskip 2pt}
A1  &00:00:42.0&-01:26:40.2&$24.91\pm0.20$&$23.39\pm0.15$&$21.94\pm0.04$&$21.34\pm0.04$&$20.92\pm0.04$&$20.25\pm0.04$&$19.99\pm0.04$&1.677\cr
A2  &00:00:42.4&-01:26:26.7&$24.67\pm0.15$&$23.78\pm0.18$&$22.08\pm0.05$&$21.33\pm0.04$&$20.88\pm0.04$&$20.13\pm0.03$&$19.78\pm0.03$&1.684\cr
B2a &00:00:42.1&-01:25:32.1&$27.15\pm1.44$&$\dots$&$22.71\pm0.09$&$22.36\pm0.09$&$22.35\pm0.15$&$21.54\pm0.05$&$21.49\pm0.08$&1.681\cr
\noalign{\vskip 5pt\hrule\vskip 2pt}
A3  &00:00:41.8&-01:26:21.5&$27.44\pm2.00$&$24.01\pm0.23$&$21.76\pm0.04$&$21.28\pm0.04$&$21.14\pm0.05$&$20.69\pm0.04$&$20.54\pm0.04$&2.019\cr
A4a &00:00:42.8&-01:26:27.0&$24.45\pm0.13$&$23.67\pm0.17$&$21.98\pm0.05$&$21.44\pm0.04$&$21.08\pm0.05$&$20.32\pm0.04$&$20.11\pm0.04$&2.032\cr
B2c$^{a}$&00:00:41.9&-01:25:29.3&$25.43\pm0.30$&$24.41\pm0.33$&$22.51\pm0.07$&$21.98\pm0.06$&$21.70\pm0.08$&$20.69\pm0.04$&$20.39\pm0.04$&2.029\cr
C1a$^{b}$ &00:00:44.8&-01:26:17.7&$24.75\pm0.16$&$23.16\pm0.11$&$21.23\pm0.03$&$20.57\pm0.02$&$20.08\pm0.03$&$19.25\pm0.03$&$18.80\pm0.03$&2.039\cr
C1b$^{b}$ &00:00:44.8&-01:26:16.7&$24.75\pm0.16$&$23.16\pm0.11$&$21.23\pm0.03$&$20.57\pm0.02$&$20.08\pm0.03$&$19.25\pm0.03$&$18.80\pm0.03$&2.039\cr
D1  &00:00:45.0&-01:25:36.5&$25.56\pm0.36$&$24.53\pm0.40$&$22.46\pm0.07$&$21.78\pm0.06$&$21.41\pm0.07$&$20.48\pm0.04$&$20.31\pm0.04$&2.018\cr
\noalign{\vskip 5pt\hrule\vskip 2pt}
A4c &00:00:43.1&-01:26:31.3&$22.48\pm0.03$&$20.97\pm0.02$&$20.43\pm0.02$&$20.09\pm0.02$&$19.88\pm0.02$&$20.08\pm0.03$&$20.49\pm0.04$&1.615\cr
A5 &00:00:41.9&-01:26:13.1&$24.85\pm0.18$&$24.58\pm0.40$&$23.50\pm0.17$&$22.20\pm0.08$&$22.33\pm0.13$&$21.47\pm0.05$&$21.36\pm0.06$&2.121\cr
B2b$^{a}$&00:00:41.9&-01:25:29.3&$25.43\pm0.30$&$24.41\pm0.33$&$22.51\pm0.07$&$21.98\pm0.06$&$21.70\pm0.08$&$20.69\pm0.04$&$20.39\pm0.04$&1.921\cr
B3  &00:00:43.0&-01:26:00.3&$24.12\pm0.09$&$23.47\pm0.13$&$22.02\pm0.05$&$21.56\pm0.04$&$21.23\pm0.05$&$20.68\pm0.04$&$20.45\pm0.04$&1.976\cr
\noalign{\vskip 3pt\hrule\vskip 3pt}
}}
\endPlancktablewide
\justify
\end{table*}

Figure~\ref{fig:twodspec} shows the 2-D spectra of the sources found at the locations of
the two structures, presenting the velocity offset with respect to a fixed reference frame at $z=2.03$
and $z=1.68$, to account for the two high-$z$ objects.
At $z\simeq2.03$, we detect $H_{\alpha}$ emission in a clear manner for D1 and C1b,
and, less clearly for A4a. B2c, A4 and A3 have a faint $H_{\alpha}$  line which can be identified on the 2D spectral image. C1a shows a broad emission line (more than 200 \AA\ on the spectrum) and the $H_{\alpha}$ identification is confirmed by the detection of the [\ion{S}{II}]$\lambda$6717-6731 doublet. 
At $z\simeq1.68$, the emission from $H_{\alpha}$ is strong or clearly detected for A1, A2 and B2a.

For several galaxies, we see other lines in their spectra,
including the aforementioned [\ion{N}{II}]$\lambda$6583 (A1, D1) and [\ion{O}{III}]$\lambda$5007 lines
(A1, B2a, C1b, A4a) that are typical in galaxies with high star-formation rates.
However, given the intrinsic faintness of our sources, most of these lines are weak and/or noisy,
and are simply used for consistency check of the $H_{\alpha}$ identification. They cannot be used to further constrain the physical properties of the galaxies.

We have fitted the 1-D spectra of the galaxies associated with either structure
to obtain more accurate redshift estimates.
For the fit, we have adopted an educated guess of the redshift (arising from the 2-D spectrum), assumed a Gaussian profile for the $H_{\alpha}$ line plus a continuum and added,
where necessary, the presence of emission coming from the [\ion{N}{II}]$\lambda$6583 line.
This fit also allows us to estimate the flux of the $H_{\alpha}$ line,
which, as discussed in Sect.~\ref{sec:discussion} below,
is used to derive the $H_{\alpha}$ luminosity ($L_{H\alpha}$) for each
of these galaxies. The measured $H_{\alpha}$ flux should be corrected from
slit-loss, as only a fraction of the source flux enters the X-shooter
instrument. It is very difficult to estimate this factor
for each galaxy because it strongly depends on the slit position with respect
to the centroid of the source, on the seeing during the observations and on the intrinsic distribution of $H_{\alpha}$ emission and extinction within the galaxy. An
estimate of this effect is proposed by \citet{kruhler15} from a large sample
of observations of GRB hosts with X-shooter: their median slit-loss factor is
$1.6 ^{+0.7}_{-0.5}$. We decided to apply a correction factor of 1.6 to all our measures.   
The results are summarised in Table~\ref{table:fit_results}.

\begin{table}
\centering
\caption{Redshifts and SNR of the $H_{\alpha}$ line of the spectroscopically observed galaxies associated with the structures at $z\simeq1.7$ and $z\simeq2.0$.
Redshift uncertainties are dominated by the wavelength calibration uncertainty (equivalent to $\pm 0.0008$). 
The SNR values are derived from the integrated flux of the line and its uncertainty.
From spectral fits, we have then obtained $H_{\alpha}$ line fluxes and  
luminosities, corrected from slit-loss but without any correction for extinction. 
For C1a, due to a broad emission line, we have large uncertainties
in the fit of the $H_{\alpha}$ emission. }
\label{table:fit_results}
\nointerlineskip
\setbox\tablebox=\vbox{
\newdimen\digitwidth % These five lines change what an asterisk
\setbox0=\hbox{\rm 0} % means to TeX. Instead of meaning
\digitwidth=\wd0 % "print an '*' here", it now means "leave
\catcode`*=\active % as much blank space as a single number
\def*{\kern\digitwidth} % takes up".
\newdimen\signwidth % These five lines change the meaning of an
\setbox0=\hbox{{\rm +}} % exclamation mark in the same way, so that it
\signwidth=\wd0 % leaves as much space as a plus or minus sign.
\catcode`!=\active % These definitions will disappear at the end of
\def!{\kern\signwidth} % the \vbox.
\newdimen\pointwidth % These five lines change the meaning of a
\setbox0=\hbox{.} % question mark in the same way, so that it
\pointwidth=\wd0 % leaves as much space as a decimal point.
\catcode`?=\active % These definitions will disappear at the end of
\def?{\kern\pointwidth} % the \vbox.
\halign{
#\hfil\tabskip=1.0em&
\hfil#\hfil\tabskip=1.0em&
\hfil#\hfil\tabskip=1.0em&
\hfil#\hfil\tabskip=1.0em&
\hfil#\hfil\tabskip=0pt\cr
\noalign{\doubleline}
ID&$z_{spec}$&SNR&$f_{H\alpha}$&$L_{H\alpha}$\cr
 & & & $[10^{-17} {\rm erg/s/cm}^2]$ & $[10^{41} {\rm erg/s}]$\cr
\noalign{\vskip 5pt\hrule\vskip 2pt}
A1  &1.6777&16&$7.8\pm0.5$&$16.3\pm1.1$\cr
A2  &1.6839&*7&$4.6\pm0.6$&$*9.8\pm1.1$\cr
B2a&1.6819&*4&$2.9\pm0.6$&$*5.9\pm1.3$\cr
\noalign{\vskip 5pt\hrule\vskip 2pt}
A4a &2.0383&*2&$1.8\pm0.8$&$*5.9\pm2.9$\cr
C1a&2.0311&*4&$29.1\pm7.0$&$96.1\pm23.4$\cr
C1b&2.0387&14&$11.4\pm0.8$&$37.9\pm2.4$\cr
D1&2.0186&*3&$2.7\pm0.8$&$*9.0\pm2.9$\cr
\noalign{\vskip 3pt\hrule\vskip 3pt}
}}
\endPlancktable
\justify
\end{table}

% DISCUSSION
\section{Discussion}
\label{sec:discussion}

Our analysis reveals an over-density of high-$z$ galaxies,
detected using optical/NIR colours and confirmed spectroscopically,
at the location of the \Planck\ high-$z$ candidate source PHz\,G95.5$-$61.6. 
All the indicators used lead to the same global picture: \Planck\ detected the combined emission
of high-$z$ galaxies (in the redshift range between 1.7 and 2.0)
within a $1^{\prime}$-distance from this source, mostly associated with two structures.

\subsection{3D spatial distribution of the galaxies}
\begin{figure*}
\center
\includegraphics[width=13.0cm]{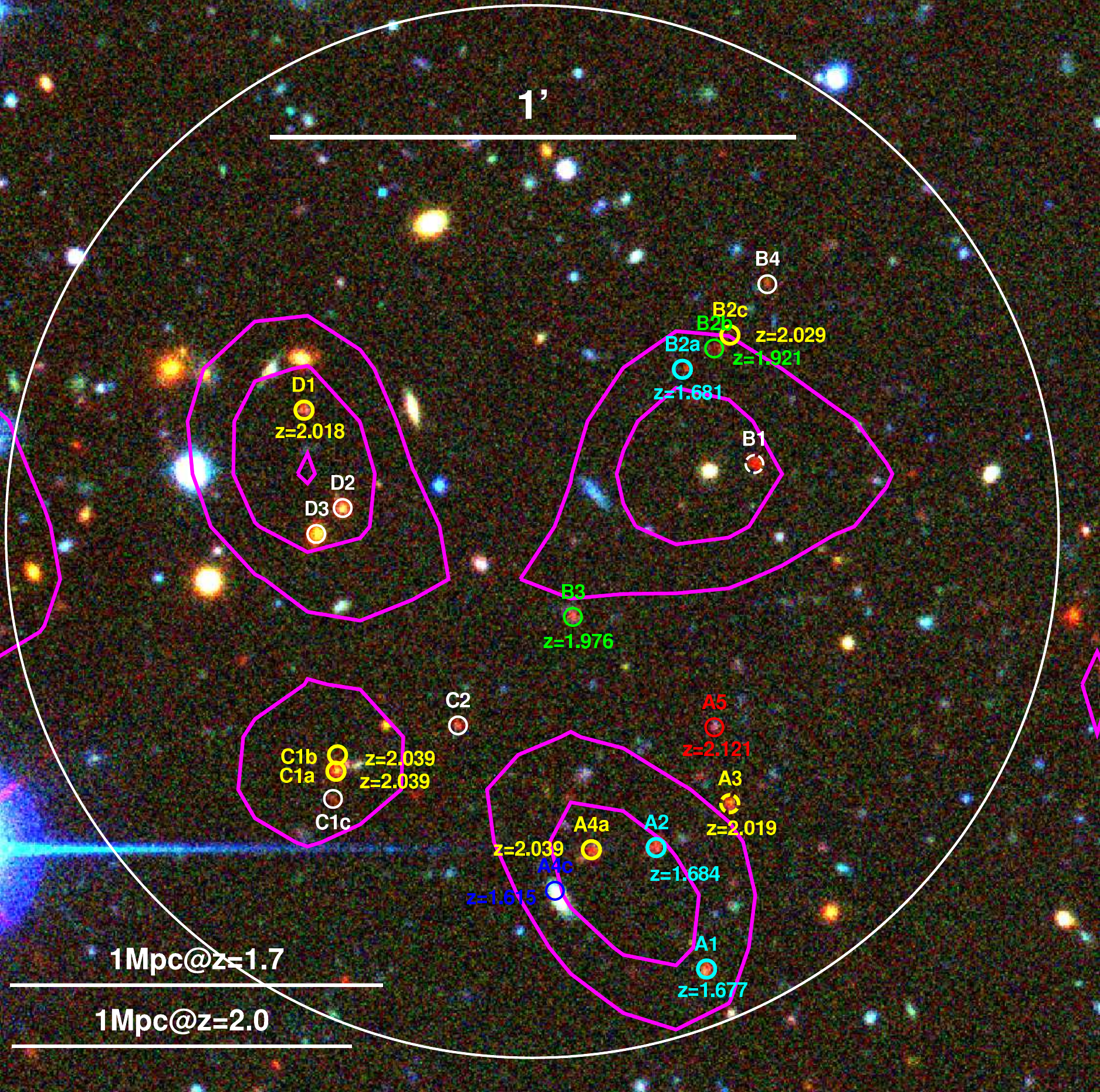} 
\caption{Zoom-in from Fig.~\ref{fig:false_colour_CFHT} showing the location of  the spectroscopically observed galaxies. The contours of the 250\,\micron\ {\it Herschel}/SPIRE map are over-plotted in magenta.
These galaxies have been colour-coded according to their spectroscopic redshifts (also included in the figure), from dark blue (closest galaxy, i.e., A4c, $z=1.615$) to red (farthest, A5, $z=2.121$).
The members of the two high-$z$ structures are then shown in cyan ($z\simeq1.7$) and yellow ($z\simeq2.0$).
The galaxies for which we could not determine their spectroscopic redshifts (B1, B4, C2, D2 and D3) are marked in white.
For reference, a circle of 1\arcmin\ radius centred at ${\rm RA}={\rm 00:00:43.31}$ and ${\rm Dec}=-{\rm 01:25:50.5}$ is shown in white.
In addition, the scales corresponding to a comoving size of 1\,Mpc at $z\simeq1.7$ and $z\simeq2.0$ are also shown in the bottom left corner.
North is up and East is left.}
\label{fig:summary}
\end{figure*}

Based on the redshift distributions we obtained, the two structures are not physically connected.
It should be noted here that there is a gap in the atmospheric window in the NIR, 
where the atmospheric transmission is very low.
This makes it difficult to spectroscopically confirm the presence of galaxies at $1.8\le z\le2.0$,
and could, in principle, bias a spectroscopic study toward a double-peaked redshift distribution irrespective of the intrinsic redshift distribution.
However, since we do detect two galaxies at redshifts intermediate between our two high-$z$ structures and since the lower redshift structure is at $z\simeq1.7$ (well below the cut-off set by the atmospheric transmission) we are confident that we indeed detect two separate structures
aligned along the line of sight, interpreted as two collapsing nodes across the Cosmic web. 
On the other hand, the galaxy members of each node do seem to be physically linked to each other
and belong to the same potential well, as, in both cases, the galaxies are all encompassed within 
a comoving radius of $\lesssim1.0\,{\rm Mpc}$ (see Fig.~\ref{fig:summary}).
This size is compatible with the virial sizes of local clusters and recently discovered proto-clusters
at $z>1.5$ \citep[e.g.,][]{Castellano2007, Andreon2010, Andreon2011, Gobat2013}.

The fact that most of the {\it Herschel} sources associated with PHz\,G95.5$-$61.6
do not spatially coincide with any over-density of colour-selected galaxies in a strict sense,
but are located in its outskirts, is not disturbing. 
In fact, studies of X-ray selected clusters at low, intermediate and high redshifts have shown that
these evolved, massive DM halos can host a population of FIR/sub-mm bright sources
at the boundaries of their virialized regions \citep{Braglia2011, Santos2013}
or along their matter-feeding filaments \citep{Biviano2011}.
Furthermore, there is growing evidence that infalling galaxies can experience significant episodes
of star-formation activity on the outskirts of clusters and, in general, along filamentary structures
up to $z \simeq 1$ \citep[e.g., ][]{Porter2007, Braglia2007, Porter2008, Darvish2014}.
A physical explanation for this phenomenon has been recently given by \citet{Mahajan2012}:
according to these authors, ``a relatively high galaxy density in the infall regions of clusters
promotes interactions amongst galaxies, leading to momentary bursts of star formation''.
The presence of a few sub-mm bright sources (each one potentially associated with just one galaxy
or with a few galaxies at most) at the boundaries of the galaxy distribution in PHz\,G95.5$-$61.6
is fully consistent with the previous observational results and this new astrophysical scenario in galaxy evolution.

\subsection{Star formation rate and stellar mass}
\begin{table*}
\centering
\caption{Values for the SFR and the stellar mass obtained from direct
measurement on the observed spectra and photometry of each galaxy. The specific star formation rate (sSFR) is
defined as the ratio between these two quantities. The corrected SFR is
deduced by applying the extinction factor $A (H_{\alpha})$ line, with the relation measured
in \citet{kashino13}. C1a has been removed from the table because if is identified with a extremely red
QSO so these values do not make sense. C1b is blended with C1a
and $K_s$ photometry is not reliable.
}
\label{table:sfr_results}
\nointerlineskip
\setbox\tablebox=\vbox{
\newdimen\digitwidth % These five lines change what an asterisk
\setbox0=\hbox{\rm 0} % means to TeX. Instead of meaning
\digitwidth=\wd0 % "print an '*' here", it now means "leave
\catcode`*=\active % as much blank space as a single number
\def*{\kern\digitwidth} % takes up".
\newdimen\signwidth % These five lines change the meaning of an
\setbox0=\hbox{{\rm +}} % exclamation mark in the same way, so that it
\signwidth=\wd0 % leaves as much space as a plus or minus sign.
\catcode`!=\active % These definitions will disappear at the end of
\def!{\kern\signwidth} % the \vbox.
\newdimen\pointwidth % These five lines change the meaning of a
\setbox0=\hbox{.} % question mark in the same way, so that it
\pointwidth=\wd0 % leaves as much space as a decimal point.
\catcode`?=\active % These definitions will disappear at the end of
\def?{\kern\pointwidth} % the \vbox.
\halign{
#\hfil\tabskip=1.0em&
\hfil#\hfil\tabskip=1.0em&
\hfil#\hfil\tabskip=1.0em&
\hfil#\hfil\tabskip=1.0em&
\hfil#\hfil\tabskip=1.0em&
\hfil#\hfil\tabskip=1.0em&
\hfil#\hfil\tabskip=0pt\cr
\noalign{\doubleline}
ID&$z_{spec}$&SFR&$M_{\star}^{K}$&sSFR($H_\alpha$)&$A (H_\alpha$)&SFR$_{\rm
corr}$\cr
 & & $[\Msun\,{\rm yr}^{-1}]$&$[\Msun]$&$[{\rm yr}^{-1}]$ & mag.&$[\Msun\,{\rm yr}^{-1}]$\cr
\noalign{\vskip 5pt\hrule\vskip 2pt}
A1 &1.6777&$13.0\pm1.0$&$1.7\times10^{11}$&$7.6\times10^{-11}$&$2.0$&$82\pm6$\cr
A2 &1.6839&$*7.8\pm1.0$&$1.8\times10^{11}$&$4.3\times10^{-11}$&$2.0$&$49\pm6$\cr
B2a&1.6819&$*4.6\pm1.0$&$0.5\times10^{11}$&$9.2\times10^{-11}$&$1.4$&$17\pm4$\cr
\noalign{\vskip 5pt\hrule\vskip 2pt}
A4a&2.0383&$*4.8\pm2.2$&$3.0\times10^{11}$&$1.6\times10^{-11}$&$2.3$&$40\pm18$\cr
C1b&2.0387&$30.1\pm1.9$& & & & \cr
D1&2.0186&$*7.2\pm2.2$&$1.1\times10^{11}$&$6.5\times10^{-11}$&$1.8$&$38\pm11$\cr
\noalign{\vskip 3pt\hrule\vskip 3pt}
}}
\endPlancktablewide
\justify
\end{table*}

From the fitted $H_{\alpha}$ line fluxes
 and the measured redshifts, 
we have computed the $H_{\alpha}$ luminosity ($L_{H\alpha}$)
and hence the SFR estimates for the galaxies that are confirmed members of the two proto-clusters,
following \citet{Kennicutt1998}. 
Our results are summarised in Table~\ref{table:sfr_results}. We also computed
the stellar masses of these galaxies,using the relation from \citet{daddi04}
that relies on the Ks-band magnitude. We adopted the normalisation for the SED method used by the
latter authors and a Salpeter (1955) IMF. 
But for most of the high-$z$ star forming galaxies it is currently admitted
that dust extinction is an important factor that has a major influence on
the measured  $H_{\alpha}$ fluxes. To evaluate this attenuation and to
correct it, we followed the results obtained by \citet{kashino13,kashino14} on a sample
of high redshift star forming galaxies (sBzK galaxies at $1.4<z<1.7$)
detected in the COSMOS field \citep{mccracken10}. A correlation is found between the
$H_{\alpha}$ attenuation, measured spectroscopically from the
$H_{\alpha}/H_{\beta}$ line ratio, and the stellar mass obtained from the
BzK photometry. We used their best fit equation and applied it to our data to
obtaind a corrected SFR for the few galaxies for which data is available. This
correcting factor is on average about 6. 
The majority of our galaxies shows SFRs of a few tens of $\Msun\,{\rm
yr}^{-1}$. Our estimations put all
these galaxies below the ``main sequence'' part of the SFR vs stellar mass diagram
\citep[see Fig.~1 of ][]{Rodighiero2011}, in agreement with the
moderate optical colour we see. 
Hence, these galaxies represent average star forming galaxies at $z\sim2$.

\begin{table}
\centering
\caption{Dust temperature, FIR luminosities and FIR-based SFRs for the {\it Herschel}/SPIRE sources within the \Planck\ beam.
We provide two estimations for each of these quantities assuming that all the FIR signal 
arises from sources either at $z=1.7$ (top part of the table) or at $z=2.0$ (bottom),
which serve as lower and upper limits (respectively) to the expected FIR estimations.}
\label{table:IR_SFR}
\nointerlineskip
\setbox\tablebox=\vbox{
\newdimen\digitwidth % These five lines change what an asterisk
\setbox0=\hbox{\rm 0} % means to TeX. Instead of meaning
\digitwidth=\wd0 % "print an '*' here", it now means "leave
\catcode`*=\active % as much blank space as a single number
\def*{\kern\digitwidth} % takes up".
\newdimen\signwidth % These five lines change the meaning of an
\setbox0=\hbox{{\rm +}} % exclamation mark in the same way, so that it
\signwidth=\wd0 % leaves as much space as a plus or minus sign.
\catcode`!=\active % These definitions will disappear at the end of
\def!{\kern\signwidth} % the \vbox.
\newdimen\pointwidth % These five lines change the meaning of a
\setbox0=\hbox{.} % question mark in the same way, so that it
\pointwidth=\wd0 % leaves as much space as a decimal point.
\catcode`?=\active % These definitions will disappear at the end of
\def?{\kern\pointwidth} % the \vbox.
\halign{
#\hfil\tabskip=1.0em&
\hfil#\hfil\tabskip=1.0em&
\hfil#\hfil\tabskip=1.0em&
\hfil#\hfil\tabskip=1.0em&
\hfil#\hfil\tabskip=0pt\cr
\noalign{\doubleline}
ID&$T_{\rm dust}$&$L_{\rm FIR}$ &SFR \cr
 & $[K]$&$[10^{12}\Lsun]$&$[\Msun\,{\rm yr}^{-1}]$\cr
\noalign{\vskip 5pt\hrule\vskip 2pt}
A ($z=1.7$) &$25.9\pm2.9$&$2.8\pm0.4$&$482\pm*69$\cr
B ($z=1.7$) &$26.3\pm3.0$&$2.7\pm0.4$&$465\pm*69$\cr
C ($z=1.7$) &$32.0\pm7.8$&$2.2\pm0.8$&$379\pm138$\cr
D ($z=1.7$) &$31.7\pm7.0$&$2.7\pm0.8$&$465\pm138$\cr
E ($z=1.7$) &$26.4\pm5.5$&$2.1\pm0.5$&$362\pm*86$\cr
\noalign{\vskip 5pt\hrule\vskip 2pt}
A ($z=2.0$) &$29.5\pm3.5$&$4.3\pm0.6$&$740\pm103$\cr
B ($z=2.0$) &$29.5\pm3.4$&$4.1\pm0.5$&$706\pm*86$\cr
C ($z=2.0$) &$36.3\pm9.0$&$3.3\pm1.5$&$568\pm258$\cr
D ($z=2.0$) &$35.9\pm8.2$&$4.0\pm1.3$&$689\pm223$\cr
E ($z=2.0$) &$30.1\pm6.7$&$3.1\pm0.7$&$534\pm121$\cr
\noalign{\vskip 3pt\hrule\vskip 3pt}
}}
\endPlancktable
\end{table}

For an independent SFR estimation, we have used the {\it Herschel}/SPIRE data,
following the same technique described in detail in \citet{Canameras2015}.
We have fitted the data to a single-component modified blackbody SED (in an optically thick case scenario),
obtaining a dust temperature ($T_{\rm dust}$) for each source (with a fixed emissivity index $\beta=1.6$).
We have then obtained the infrared luminosities by integrating over this best-fit SED,
between 8 and 1000\,\microns\ (rest frame), and derived the SFR of each source,
following \citet{Kennicutt1998}.
This process is done assuming that all the SPIRE signal arises from galaxies belonging only to one
of the two subsystems (either at $z=1.7$ or at $z=2.0$), which yields upper and lower limits
on the FIR estimates of SFR.
The ensuing results are listed in  Table~\ref{table:IR_SFR}.

The integrated FIR-based SFR in the \Planck\ beam is equal to $\sim2000-3000\,\Msun\,{\rm yr}^{-1}$.
This estimate is of same order of magnitude as the global SFR of $4924\Msun\,{\rm yr}^{-1}$
computed by \citet{Clements2014} for a proto-cluster at $z=2.05$.

\begin{table*}
\centering
\caption{MAGPHYS $1\sigma$ confidence interval for the SFR, stellar mass, bolometric dust luminosity and sSFR derived from fitting the optical/NIR SEDs of the spectroscopic targets with the best secure redshift identification. 
These quantities have been obtained using the \citet{Chabrier2003} initial mass function (IMF). To rescale them to a \citet{Salpeter1955} IMF, an average factor of 1.4 should be applied to the SFRs and stellar masses. 
}
\label{table:SED_fit_results}
\nointerlineskip
\setbox\tablebox=\vbox{
\newdimen\digitwidth % These five lines change what an asterisk
\setbox0=\hbox{\rm 0} % means to TeX. Instead of meaning
\digitwidth=\wd0 % "print an '*' here", it now means "leave
\catcode`*=\active % as much blank space as a single number
\def*{\kern\digitwidth} % takes up".
\newdimen\signwidth % These five lines change the meaning of an
\setbox0=\hbox{{\rm +}} % exclamation mark in the same way, so that it
\signwidth=\wd0 % leaves as much space as a plus or minus sign.
\catcode`!=\active % These definitions will disappear at the end of
\def!{\kern\signwidth} % the \vbox.
\newdimen\pointwidth % These five lines change the meaning of a
\setbox0=\hbox{.} % question mark in the same way, so that it
\pointwidth=\wd0 % leaves as much space as a decimal point.
\catcode`?=\active % These definitions will disappear at the end of
\def?{\kern\pointwidth} % the \vbox.
\halign{
#\hfil\tabskip=1.0em&
\hfil#\hfil\tabskip=1.0em&
\hfil#\hfil\tabskip=1.0em&
\hfil#\hfil\tabskip=1.0em&
\hfil#\hfil\tabskip=1.0em&
\hfil#\hfil\tabskip=0pt\cr
\noalign{\doubleline}
ID&SFR&$L_{\rm dust}$&$M_{\star}$&sSFR\cr
 &$[\Msun\,{\rm yr}^{-1}]$&$[\Lsun]$&$[\Msun]$&$[{\rm yr}^{-1}]$\cr
\noalign{\vskip 5pt\hrule\vskip 2pt}
A1&$23.3-571$	& $(0.34-6.02)\times10^{12}$ 	& $(0.83-1.29)\times10^{11}$ 	      & $(0.19-6.68)\times10^{-9*}$\cr
A2&$15.5-*50$	& $(2.0*-7.1*)\times10^{11}$ 	& $(1.26-1.74)\times10^{11}$ 	      & $(1.1*-3.3*)\times10^{-10}$ \cr
B2a&$1.2-2.8$	& $(1.0*-4.0*)\times10^{10}$ 	& $(1.8*-2.3*)\times10^{10}$ 	      & $(0.7*-1.5*)\times10^{-10}$ \cr
\noalign{\vskip 5pt\hrule\vskip 2pt}
A3&$2.4-12.9$	& $(0.7*-3.1*)\times10^{11}$ 	& $(6.8*-7.91)\times10^{10}$ 	      & $(0.3*-1.7*)\times10^{-10}$ \cr
A4a&$45.4-115.3$ & $(0.56-1.48)\times10^{12}$ 	& $(1.10-1.35)\times10^{11}$ 	      & $(0.37-1.06)\times10^{-9*}$ \cr
D1&$*18.7-*98.2$	& $(0.28-1.35)\times10^{12}$ 	& $(0.98-1.41)\times10^{11}$ 	      & $(1.5*-8.4*)\times10^{-10}$ \cr
\noalign{\vskip 3pt\hrule\vskip 3pt}
}}
\endPlancktablewide
\end{table*}

We have also fitted the broad-band SEDs of the galaxies followed up with VLT/X-Shooter
to constrain their stellar masses, SFRs and bolometric dust luminosities. 
We have used the publicly available model package MAGPHYS\footnote{MAGPHYS is maintained by E. da Cunha and S. Charlot. It is available at: \url{http://www.iap.fr/magphys/magphys/MAGPHYS.html}.} 
(``Multi-wavelength Analysis of Galaxy Physical Properties``) that builds on the works by \citet{bruzual03, charlot00}. 
This code works in a two step process \citep[see ][ for a full description]{dacunha08}:
1) a library of model SEDs is assembled for a wide range of physical parameters
(characterising both the stellar and interstellar components) and the redshifts and broad-band filters
of the observed galaxies;
and 2) a marginalised likelihood distribution is built for each physical parameter of each galaxy
through comparison of the observed SED with all the synthetic ones in the library. 

From this process we obtain likelihood estimations for the SFR,
stellar mass, and dust luminosity.
The results are listed in Table~\ref{table:SED_fit_results}.
We see that the SED-fitting estimates for the stellar mass (once rescaled from a \citet{Chabrier2003}
initial mass function -- IMF -- to a \citet{Salpeter1955} IMF) are consistent with
the estimation derived from the the Ks luminosity, with sligthly lower values.
The SFR derived from SED-fitting are consistent with those derived from the
$H_{\alpha}$ flux, provided the dust attenuation correction if taken into
account. Both values are still a factor 3 to 10 below the results obtained from
the SPIRE data (Table~\ref{table:IR_SFR}) and are not easy to reconcile,
although SFR derived from MAGPHYS analysis  display a wide range of
accepted values for some galaxies. We also remind that the SFR values
obtained from the SPIRE data strongly depend on the redshift of the sources.
Moreover, based on the total SFRs of each SPIRE source (see Table~\ref{table:IR_SFR}),
we estimate that these sub-mm sources should have between one and three galaxy counterparts.
This is consistent with the sky surface density and median SFR value of $BzK$-selected star forming galaxies
\citep{daddi04} with $1.5 \leq z \leq 2.5$ in the field that we retrieve from \citet{Oteo2014}.

Finally, the stellar masses, SFRs and bolometric dust luminosities listed in Table~\ref{table:SED_fit_results} suggest
that some of galaxies under study could be classified as luminous/ultra-luminous infrared galaxies \citep[LIRGs/ULIRGs,][]{daddi05}\footnote{By definition, LIRGs and ULIRGs exhibit bolometric dust luminosities (from 8 to $1000\,\mathrm{\mu m}$) above $10^{11}$ and $[1 - 2] \times 10^{12}\,\mathrm{L_{sun}}$, respectively.}, and, thus, are likely responsible for a significant fraction of the fluxes
of the individual SPIRE sources.
In particular, the two galaxies A1 and A4a are candidate counterparts to SPIRE source A.
Furthermore, the close pair C1a and C1b is a strong candidate for being associated with SPIRE source C, especially the red quasar C1a. 
We note that the galaxies A1, A4a, C1a, C1b and D1 are both candidate ULIRGs
and spectroscopic members of either galaxy structure associated with PHz\,G95.5$-$61.6.

% CONCLUSION
\section{Conclusions}
\label{sec:conclusions}

We combined the {\Planck}/HFI and {\it IRAS} all-sky maps to identify the brightest, and intrinsically rarest,
sub-mm emitters in the distant Universe over the cleanest 35\% of the sky.
We used colour-colour criteria plus flux density and signal-to-noise threshold
to build a sample of several hundreds of {\Planck} high-redshift candidates,
amongst which we found PHz\,G95.5$-$61.6.
We presented and discussed here the full characterisation of this candidate,
used as the bench-mark for an ongoing follow-up programme on the full catalogue.

Our analysis shows that the excess signal detected by \Planck\ for PHz\,G95.5$-$61.6
is indeed correlated with an over-density of ``red sources'' as seen by {\it Herschel}.
The analysis of the optical and IR data from CFHT and {\it Spitzer} shows that
the IR excess seen by SPIRE is caused by an excess of galaxies with typical colours
of $i-K_{\rm s}\sim2.5$ and $J-K_{\rm s}\sim1.0$, consistent with being dusty, star forming galaxies at $z\sim2$.
Photometric and spectroscopic redshifts of the galaxies in the same region of the skyfurther support that
the \Planck\ detection is caused by an overabundance of high-$z$ galaxies. 
However, these galaxies are not members of a single system, but rather of two galaxy clumps
viewed along the same line of sight, and located at  $z\simeq1.7$ and $z\simeq2.0$.

The physical size ($\sim1$\,Mpc)
of our two proto-cluster candidates is consistent with known proto-clusters/groups at $1\le z\le 2$. 
The star-formation rates (SFRs) of the spectroscopic members suggest that
our proto-clusters host intense star formation activity, consistent with the models that predict a peak
in the cosmic SFR density at $z\sim2$.
  
Thus, PHz\,G95.5$-$61.6 has not only proven \Planck's potential for detecting high-$z$ objects
but also that the observational strategy we have applied for the follow-up of these sub-mm sources
is appropriate for understanding their nature.

Moreover, our results show that PLCKHZ G95.5$-$61.6 is a unique and extremely interesting
alignment of two high-$z$ proto-clusters along the line of sight (not physically linked to each other),
which host an extremely vigorous star formation activity.

%BIBLIO
\bibliographystyle{aa}
%\bibliography{biblio_v1.5,Planck_bib_Oct14}

\begin{acknowledgements}
We are greatly indebted to the staff and directors of CFHT, and VLT, for their generous allocation of observing time for 
PHz\,G95.5$-$61.6 and spectacular help in carrying out or observations promptly and with great success.
We would like to thank D. Scott, N. Welikala, M. B{\'e}thermin, M. Limousin and M. Polletta for their helpful input discussing this work.
We would like to thank the anonymous referee for useful comments and suggestions that helped to 
improve the quality of this paper.

% Planck
A description of the Planck Collaboration and a list of its members can be found at
 \href{http://www.rssd.esa.int/index.php?project=PLANCK\&page=Planck\_Collaboration}{the ESA website.}
The development of {\it Planck} has been supported by: ESA; CNES and \linebreak CNRS/INSU-IN2P3-INP (France); 
ASI, CNR, and INAF (Italy); NASA and DoE (USA); STFC and UKSA (UK); CSIC, MICINN and JA (Spain); Tekes, AoF 
and CSC (Finland); DLR and MPG (Germany); CSA (Canada); DTU Space (Denmark); SER/SSO \linebreak 
(Switzerland); RCN (Norway); SFI (Ireland); FCT/MCTES (Portugal); and DEISA (EU). 
% Herschel
The Herschel spacecraft was designed, built, tested, and launched under a contract to ESA managed by the 
{\it Herschel/Planck} Project team by an industrial consortium under the overall responsibility of the prime 
contractor Thales Alenia Space (Cannes), and including Astrium (Friedrichshafen) responsible for the payload 
module and for system testing at spacecraft level, Thales Alenia Space (Turin) responsible for the service module, 
and Astrium (Toulouse) responsible for the telescope, with in excess of a hundred subcontractors.
% Spitzer
This work is based in part on observations made with the Spitzer Space Telescope, which is operated 
by the Jet Propulsion Laboratory, California Institute of Technology under a contract with NASA.
% CFHT
Based on observations obtained at the Canada-France-Hawaii Telescope (CFHT) which is operated by the 
National Research Council of Canada, the Institut National des Sciences de l'Univers of the Centre National 
de la Recherche Scientifique of France, and the University of Hawaii.
% ESO VLT
Based on observations made with ESO Telescopes at the La Silla Paranal Observatory under programme IDs 
287.A-5063 and 089.A-0276. 

% ANR Multiverse
IFC, LM and EP acknowledge the support from grant ANR-11-BS56-015.
% Labex
DP acknowledges the financial support from Labex OCEVU.
This work has been carried out thanks to the support of the OCEVU Labex (ANR-11-LABX-0060) and 
the A*MIDEX project (ANR-11-IDEX-0001-02) funded by the ``Investissements d'Avenir'' French 
government program managed by the ANR.
%%% Software
% HEALPIX
We acknowledge the use of the {\sc HEALPix} software \citep{gorski2005}.
\end{acknowledgements}

\end{document}